\documentclass[11pt]{article}

\usepackage{graphicx}
\usepackage{amsmath,amssymb,amsfonts,amsbsy,amsthm,graphicx}
\usepackage{thmtools}
\usepackage{fullpage}
\usepackage{subfigure}
\usepackage[round]{natbib}
\usepackage{array}
\usepackage{float}
\usepackage{booktabs}

\newtheorem{theorem}{Theorem}[section]

\declaretheorem[style=definition]{example}
\renewcommand\thmcontinues[1]{Continued}

\newtheorem{remark}[theorem]{Remark}

\newcommand{\gp}{G{\^a}rleanu and Pedersen }

\newcommand{\ltwo}{\mathcal{L}_2}

\usepackage{parskip}

\begin{document}

\title{Optimal Trading under Instantaneous and Persistent Price Impact, Predictable Returns and Multiscale Stochastic Volatility\footnote{An unpublished earlier version without price impact was circulated under the title ``Optimal Trading with Predictable Return and Stochastic Volatility".}}

\author{Patrick Chan\thanks{Formerly Program in Applied and Computational Mathematics, Princeton University,
{\em yukchan@princeton.edu}.} \and Ronnie Sircar\thanks{Department of Operations Research \& Financial Engineering, Princeton University, {\em sircar@princeton.edu}} \and Iosif Zimbidis\thanks{Department of Operations Research \& Financial Engineering, Princeton University, {\em iz2776@princeton.edu}}} 

\maketitle

\begin{abstract}
    We consider a dynamic portfolio optimization problem that incorporates predictable returns, instantaneous transaction costs, price impact, and stochastic volatility, extending the classical results of \citet{garleanu2013dynamic}, which assume constant volatility. Constructing the optimal portfolio strategy in this general setting is challenging due to the nonlinear nature of the resulting Hamilton-Jacobi-Bellman (HJB) equations. To address this, we propose a multi-scale volatility expansion that captures stochastic volatility dynamics across different time scales. Specifically, the analysis involves a singular perturbation for the fast mean-reverting volatility factor and a regular perturbation for the slow-moving factor. We also introduce an approximation for small price impact and demonstrate its numerical accuracy. We formally derive asymptotic approximations up to second order and use Monte Carlo simulations to show how incorporating these corrections improves the Profit and Loss (PnL) of the resulting portfolio strategy.
\end{abstract}


\section{Introduction}

A tractable framework for dynamic portfolio optimization under market frictions was proposed by \citet{garleanu2013dynamic, garleanu2016}, who derived closed-form trading strategies that incorporate return predictability along with both instantaneous and persistent transaction costs. Specifically, their model accounts for instantaneous costs arising from the mechanics of the limit order book, as well as persistent transaction costs, often referred to as price impact in the literature, which capture the lasting influence of trades on asset prices. Building on the literature on price impact, \citet{webster2023} provides a comprehensive and detailed overview of modern theory, offering extensive references on optimal execution and market impact, along with numerous practical insights into the implementation and real-world considerations of such models.

In this paper, we extend this line of research by introducing stochastic volatility into the dynamic trading problem. Specifically, we incorporate a multiscale stochastic volatility model governed by both slow and fast scale factors, following the framework of \citet{fouque2011multiscale}. Our paper also offers additional insights into the structure and solution of the system of equations that arises even under constant volatility, in the most general setting that includes both persistent and instantaneous transaction costs. \citet{garleanu2013dynamic} do not derive explicit solutions in this case but only state that a unique solution exists under suitable conditions. We demonstrate that when price impact is assumed to be relatively small, the system admits numerically tractable approximations. A similar asymptotic approach is employed in the recent work of \citet{Ekren2019}, where both forms of trading friction, instantaneous and persistent, are assumed to be small, leading to a solution characterized by an algebraic Riccati equation.

In contrast with the constant volatility case considered by \citet{garleanu2013dynamic}, under our stochastic volatility setting, the Hamilton-Jacobi-Bellman (HJB) equation becomes nonlinear. However, by applying the asymptotic approximation method as implemented by \citet{fouque2011multiscale}, we demonstrate that the problem can still be rendered tractable. Specifically, it can be viewed as a perturbation around the fully tractable solution obtained under constant volatility. Below, we further analyze and present the key aspects considered in constructing the framework for our dynamical optimization problem.

\begin{description}
\item[Return predictability.] One of the main aspects of a trader's work is predicting the movement of an asset to profit from those predictions. These forecasts are not limited to simple short-term projections, but often involve complex models and dynamics that incorporate mean-reversion and momentum properties.

\item[Transaction costs.] When traders execute their strategies by adjusting their positions in assets, friction arises from these changes. This friction typically comes from factors such as the bid-ask spread or execution commissions. Ideally, investors would like to follow the theoretical ``optimal'' portfolio; however, these additional costs cause a divergence from the optimal portfolio computed without accounting for such frictions. In the literature, transaction costs are sometimes referred to as instantaneous transaction costs, as they are considered not to impact the asset's price after the trade is executed.

\item[Price Impact.] Price impact is a key factor considered by traders when deciding on their strategies. It reflects the effect that trading decisions have on the asset's price. Although the first models incorporating price impact were introduced in the 1980s, their significance was quickly recognized, leading to widespread adoption in both academia and the industry. In the literature, price impact is sometimes referred to as persistent transaction costs.

\item[Stochastic volatility.] Stochastic volatility modeling plays a key role in asset pricing due to its ability to explain empirical phenomena such as the volatility smile. As a result, such models have been adapted to support more practical and accurate asset pricing frameworks. An influential paper by \citet{chacko2005dynamic} emphasizes the importance of multi-factor modeling, a concept we will build upon in this paper, as they observed that significantly different results and insights can be drawn when analyzing high-frequency versus low-frequency data.
\end{description}

In our analysis, we build on the dynamic portfolio optimization literature by incorporating all of the above features. The resulting problem remains as tractable as the one studied by \gp, even though we extend the volatility model from the simple constant case to a two-factor model with both slow and fast scale stochastic volatility components. Under our framework, explicit correction terms emerge, causing the optimal trading rate to diverge from the constant volatility case. These corrections lead to significant improvements in portfolio Profit and Loss (PnL), which we quantify using Monte Carlo simulations in the numerical section of the paper. Our contribution also emphasizes that while multiscale stochastic volatility models have been applied in frictionless problems such as Merton's portfolio optimization in the work of \citet{fouque2013portfolio}, here we adapt these ideas to a setting including market frictions, a more realistic approach.

Additionally, in a wide variety of practically relevant stochastic volatility models (e.g., Heston, exponential Ornstein–Uhlenbeck, and the 3/2-model), the correction terms to constant-volatility strategies can be derived in closed form. These correction terms yield trading strategies that align with rational and empirically observed economic behavior.

In particular, under fast-scale dynamics, traders are advised to deleverage their portfolios when the current volatility level exceeds its long-term average, regardless of the return-volatility correlation. In contrast, under slow-scale settings, the return-volatility correlation plays a more substantial role. Specifically, when the correlation between volatility and returns is positive, the investor optimally reduces their trading rate, anticipating that higher expected returns are accompanied by higher volatility. Furthermore, we show that the impact of slow-scale stochastic volatility is more pronounced than that of fast-scale volatility in the context of our infinite-horizon optimal trading problem. In fact, the leading-order correction in the fast-scale volatility expansion vanishes identically, and one must compute the second-order expansion to capture the primary effects of fast-scale volatility.

More recent work has focused on the dynamics of price impact. An empirical analysis by \citet{carmona2019ow} demonstrated that the propagator used in more sophisticated models of price impact exhibits local concavity for a general order book, providing a rigorous mathematical framework that supports the earlier numerical findings of \citet{Bouchaud2009}. In the latest developments, \citet{MuhleKarbe2024} show that introducing randomness in the impact level can effectively capture the effect of concavity, resulting in a simple yet time-varying linear model. Additionally, \citet{Brokmann2025} propose a framework in which nonlinear price impact can be approximated using linear trading strategies.

Recent research incorporates machine and deep learning methods into portfolio optimization to construct optimal trading strategies. For example, \citet{buehler2019deep} demonstrate a hedging framework for path-dependent payoffs using neural networks under the impact of trading friction, providing an extension to classical replication strategies. In more general settings, \citet{han2018solving} incorporate deep learning techniques to solve high-dimensional forward-backward stochastic differential equations usually arising in nonlinear stochastic control problems. \citet{heaton2017deep} illustrate how deep neural networks can capture the complex structure of asset interactions, allowing for an alternative approach to traditional portfolio construction.

Deep learning and robust control techniques have also been applied to address model uncertainty in portfolio optimization. \citet{pham2022portfolio} study a robust dynamic mean-variance portfolio selection problem under drift ambiguity and derive a general “separation principle” that reduces the optimal control problem to a simple parametric computation of the premium function. \citet{lin2022multiagent} propose a multiagent-based deep reinforcement learning framework for portfolio management that adjusts risk exposure in response to changing market conditions. \citet{cai2025bayesian} develop a Bayesian learning framework under a minimax rule, allowing an investor to update beliefs about the drift and make allocation decisions through the lens of model uncertainty. Additionally, \citet{MuhleKarbe2024efficient} compare the performance of \gp\ solutions to that of neural network-based techniques. In particular, they show that when the linear impact parameters are optimally calibrated to the underlying nonlinear impact model, the resulting policy achieves performance that is competitive with neural network methods, except in extremely illiquid markets.

\citet{abi2022optimal} and \citet{neuman2022optimal} further advance this literature by incorporating predictive finite-variation signals into propagator-type models with both transient and temporary price impact. While \citet{abi2022optimal} derive explicit optimal trading strategies using an infinite-dimensional stochastic control approach, \citet{neuman2022optimal} characterize signal-adaptive execution strategies through coupled forward-backward stochastic differential equations (FBSDEs).

The use of asymptotic approximation under multiscale stochastic volatility modeling has been applied in both optimal investment and derivative pricing problems. As described in the book by \citet{fouque2011multiscale}, singular and regular perturbation methods have been developed to effectively approximate linear option pricing problems.

Our analysis and approach are similar to this line of work, but we diverge in several critical aspects. The main difference is that we explicitly incorporate transaction costs and price impact, which leads to more realistic market dynamics. Moreover, we consider a mean-variance optimization problem with an infinite trading horizon, which better aligns with problems of practical interest. Finally, this paper contributes the explicit computation of terms up to the second-order correction in the fast-scale stochastic volatility.

\paragraph{Summary}
In Table~\ref{tab:literature} we summarize the models for dynamic trading in the literature.
Type refers to continuous or discrete-time model;
Cont. stands for continuous;
Disc. stands for discrete;
(g)BM stands for (geometric) Brownian motion;
SV stands for stochastic volatility;
pred. stands for predictability;
Lrr stands for Linear rebalancing rules.
\begin{table}[htb]
\caption{Dynamic trading models: problems, models and solution approaches.}
\label{tab:literature}
\centering
\resizebox{\textwidth}{!}{
\begin{tabular}{lllllll}
\toprule
&Type&Price dynamics&Trading friction&Price Impact&Objective&Solution\\
\midrule
\citet{merton1971optimum} & Cont. & gBM & None & No& Utility & Analytic\\
\citet{liu2002optimal} & Cont. & gBM & Proportional & No& Utility & Analytic\\
\citet{kraft2005optimal} & Cont. & gBM+Heston & None & No& Utility & Analytic\\
\citet{chacko2005dynamic} & Cont. & gBM+3/2 & None & No& Utility & Analytic\\
\citet{moallemi2012dynamic} & Disc. & BM+pred. & Quadratic & No& Mean-Variance & lrr\\
\citet{obizhaeva2013optimal} & Cont. & BM & None & Linear & Execution Cost & Analytic \\
\citet{bichuch2014optimal} & Cont. & gBM+SV & Proportional & No& Utility & Asymptotic\\
\citet{papanicolaou2014perturbation} & Cont. & gBM+pred. & None & No& Utility & Asymptotic\\
\citet{fouque2013portfolio} & Cont. & gBM+SV & None & No& Utility & Asymptotic\\
\citet{garleanu2013dynamic, garleanu2016} & Cont./Disc. & BM+pred. & Quadratic & Linear& Mean-Variance & Analytic\\
\citet{passerini2015optimal} & Cont. & BM+pred. & Linear-quadratic & No& Mean-Variance & Approximate\\
\citet{curato2017execution} & Cont. & BM & Locally concave & Locally concave & Execution Cost & Numerical \\
\citet{CollinDufresne2020} & Cont./Disc. & Markovian (switching model) & Quadratic & None & Mean-Variance & Analytic\\
\citet{neuman2022optimal} & Cont. & BM+pred. & None & Linear & Execution Cost & Analytic \\
\citet{abi2022optimal} & Cont. & BM+pred. & None & General propagator & Execution Cost & Analytic \\
\citet{MuhleKarbe2023dynamic} & Cont. & BM+partially pred. & Quadratic & None & Utility & Analytic \\
\bf{This paper} & \bf{Cont.} & \bf{BM+pred.+SV} & \bf{Quadratic} & Linear & \bf{Mean-Variance} & \bf{Asymptotic}\\
\bottomrule
\end{tabular}
}
\end{table}

\subsection{Organization}

In Section \ref{subsec: constant} we introduce the continuous-time model with the multiscale stochastic volatility and we derive the HJB equation for the optimal portfolio problem and give the analytical solution in the special case of constant volatility. In subsection \ref{subsec: constant} we introduce the constant volatility case analyzed by \gp and in subsection \ref{sec:approx} we propose an approximation to solve the constant case for relatively small price impact.
From this point on, to keep the presentation manageable, we focus on the analysis of the two factors separately.
We begin in Section \ref{sec:fast} with the case of fast volatility factor, which leads to a singular perturbation problem for the associated HJB equation.
In Section \ref{sec:slow}, we analyze the case of slowly volatility factor, which leads to a regular perturbation problem.
Section \ref{sec:multi} discusses how the fast and slow results can be combined for approximations under multiscale stochastic volatility.
In Section \ref{sec:numeric}, we illustrate our results with numerical examples.
Section \ref{sec:conclusion} concludes and suggests the extension directions.

\section{Market trading model}\label{sec:model}

\subsection{Market Friction}

When executing market orders, a cost arises due to the impact of trading decisions on liquidity and, consequently, on the price of the stock. In the existing literature, this cost is typically modeled as quadratic in the trading speed, as proposed by \citet{garleanu2016}. Denoting \( q_t \) as the position (number of shares held) at time \( t \), we have 
\begin{equation}
    dq_t = u_t \, dt,
\end{equation} 
so that \( u_t \) is the trading rate. This cost is usually introduced under the assumption that, after the execution of the trade, liquidity reverts to its original level; therefore, there is no lasting impact on the stock price. This cost is also referred to as the \emph{instantaneous transaction cost} or \emph{temporary transaction cost} in the literature. Hence, it is given by
\begin{equation}\label{eq:temp_tc}
    TC_{\text{instant}}(u_t) = \frac{K}{2} u_t^2,
\end{equation}
where \( K > 0 \) is a positive constant governing the level of transaction cost. The interpretation is that the transaction price of the asset exceeds the unaffected price process when \( u_t > 0 \), and the difference is proportional to the trading rate.

In addition to temporary transaction costs incurred upon trade execution, we also consider an additional cost commonly referred to as \emph{persistent} or \emph{transient} transaction cost. This reflects the effect that executed trades have on the price of the stock. In particular, buying or selling a quantity of shares can move the price up or down, respectively. This cost is typically associated with the term \emph{price impact}, as it affects the stock price even after the trade is completed, reflecting the possibility that liquidity may not fully recover instantly after execution.

To model this cost, we introduce the concepts of the \emph{unperturbed} price \(P_t\) and the \emph{perturbed} price \(S_t\), following \citet{webster2023}. Specifically,
\begin{equation*}
    S_t = P_t + l_t,
\end{equation*}
where \((l_t)_{t \geq 0}\) corresponds to the incurred price impact. Following \citet{garleanu2016} and \citet{webster2023}, we assume that the price impact evolves according to specific dynamics. In particular, it is linear in the change in portfolio position and decays exponentially over time. Thus, it can be modeled as:
\begin{equation}\label{eq:pi}
    dl_t = -\beta \, l_t \, dt + \lambda \, dq_t = (\lambda u_t - \beta l_t)\, dt,
\end{equation}
where \(\beta\) and \(\lambda\) are positive constants. \citet{webster2023} refer to \(\beta\) as the \emph{decay rate} of the price impact and \(\lambda\) as a measure of \emph{market liquidity}. He considers the possibility that the parameters \( \beta \) and \( \lambda \) vary over time, reflecting changes in market conditions. However, for simplicity, we assume these parameters remain constant throughout our analysis.

\subsection{Trader's Wealth Function}

In our model, the trader's wealth function consists of three components that influence the profit and loss over the trading horizon: the trader's position in the stock, the motive to avoid excessive risk exposure, and the transaction costs incurred during trading. For simplicity, we assume a zero interest rate. The trader follows a self-financing trading strategy to maximize the financial gain from their position in the stock:
\begin{equation}
    \Pi_s = \int_0^s q_t \, dS_t.
\end{equation}
From \eqref{eq:pi}, we compute:
\begin{equation*}
    q_t \, dS_t = q_t(dP_t + dl_t) = q_t dP_t - (\beta q_t l_t - \lambda q_t u_t) \, dt.
\end{equation*}

Secondly, we consider the risk aversion term. Although the trader aims to maximize profit, they simultaneously seek to minimize their exposure to risk.

We now specify the market dynamics for the stock price. For clarity, we consider a single asset with price \( P_t \) and a single return predictor \( x_t \).  
The dynamics of the price are given by
\begin{equation}\label{price}
    dP_t = \alpha_t\,dt + \sigma(Y_t,Z_t)\,dB_t,
\end{equation}
where \( B_t \) is a standard Brownian motion, and \(Y_t, Z_t\) are stochastic volatility factors.

Without loss of generality, we decompose the drift \( \alpha_t \), often referred to as the alpha signal in the literature, into a constant \( \bar\alpha \) and a zero-mean intraday component \( x_t \), so that \(\alpha_t := \bar\alpha + x_t\).

We model the signal \(x_t\) as an Ornstein–Uhlenbeck process:
\begin{equation}\label{eq:x_ou_process}
    dx_t = - \kappa x_t\,dt + \sqrt{\eta}\,dW^{(0)}_t,
\end{equation}
where \( W^{(0)}_t \) is independent of the Brownian motion \( B_t \) driving the stock price.

To account for risk aversion, we include a penalization term in the objective function:
\begin{equation}
    -\frac{\gamma}{2} \sigma^2(Y_t, Z_t) q_t^2 \, dt,
\end{equation}
where \(\gamma\) is the risk aversion parameter, regulating the intensity of penalization due to risk exposure.

Finally, incorporating the temporary transaction cost from equation \eqref{eq:temp_tc}, the trader's objective is to choose the dynamic trading strategy $\left(u_t\right)_{t\geq 0}$ to maximize the present value of all future expected excess returns, penalized for risk and trading costs:
\begin{align*}
    \mathbb{E}\left[\int_0^\infty e^{-\rho (s-t)}\left(q_s x_s - \beta q_s l_s - \lambda q_s u_s -\frac{\gamma}{2} \sigma^2(Y_s, Z_s) q_s^2 -\frac{K}{2} u_s^2\right)ds\right],
\end{align*}
where \(\rho > 0\) is the discount factor. Assuming \((x_t, Y_t, Z_t)\) are jointly Markovian, we define the value function
\begin{align*}
    v(q, l, x, y, z) = \sup_{u} \, \mathbb{E}_{q, l, x, y, z}\left[\int_0^\infty e^{-\rho t} \left( 
    q_t x_t - \beta q_t l_t - \lambda q_t u_t - \frac{\gamma}{2} \sigma^2(Y_t, Z_t) q_t^2 - \frac{K}{2} u_t^2 
    \right) \, dt \right],
\end{align*}
where we use the notation
\begin{align*}
    \mathbb{E}_{q, l, x, y, z}[\,\cdot\,] = \mathbb{E}[\, \cdot \,|\, q_0 = q,\, l_0 = l,\, x_0 = x,\, y_0 = y,\, z_0 = z],
\end{align*}
and \((q,l,x) \in \mathbb{R}^3\).

\subsection{Multiscale Stochastic Volatility}
We work under the multiscale stochastic volatility framework, as discussed in \citet{fouque2011multiscale, fouque2013portfolio} for option pricing and portfolio optimization, where there is one fast and one slow volatility factor.  
Here, the volatility is modeled as a function \(\sigma\) of a fast factor \(Y\) and a slow factor \(Z\): \(\sigma(Y_t, Z_t)\).  
The volatility-driving factors \((Y_t, Z_t)\) are described by:
\begin{align}\label{eq:multiscale_model_1}
    dY_t &= \frac{1}{\varepsilon} b(Y_t)\,dt + \frac{1}{\sqrt{\varepsilon}} a(Y_t) \,dW^{(1)}_t, \\
    dZ_t &= \delta c(Y_t)\,dt + \sqrt{\delta} g(Y_t) \,dW^{(2)}_t,
\label{eq:multiscale_model_2}
\end{align}
where \(\left(W^{(0)}_t, W^{(1)}_t, W^{(2)}_t\right)\) are standard Brownian motions on a filtered probability space \(\left(\Omega, \mathcal{F}, (\mathcal{F}_t)_{t \geq 0}, \mathbb{P} \right)\), with instantaneous correlations given by:
\[
d\langle W^{(0)}, W^{(i)} \rangle_t = \rho_i\,dt, \quad i = 1,2, \qquad d\langle W^{(1)}, W^{(2)} \rangle_t = \rho_{12}\,dt,
\]
where \(|\rho_1| < 1\), \(|\rho_2| < 1\), \(|\rho_{12}| < 1\), and \(1 + \rho_1 \rho_2 \rho_{12} - \rho_1^2 - \rho_2^2 - \rho_{12}^2 > 0\) to ensure the positive definiteness of the covariance matrix of the three Brownian motions.

We also assume that the Brownian motions \(\left(W^{(0)}_t, W^{(1)}_t, W^{(2)}_t\right)\) are independent of \(B_t\).  
The model is specified by the coefficients \(\bar\alpha, \kappa, \eta, \sigma, a, b, c, g\).  
The parameters \(\varepsilon\) and \(\delta\), when small, characterize the fast and slow variation of the volatility factors \(Y\) and \(Z\), respectively.

We assume that \(Y_t \overset{d}{=} Y^{(1)}_{t/\varepsilon}\), where \(Y^{(1)}\) is an ergodic process with a unique invariant distribution \(\Phi\), independent of \(\varepsilon\).  
Similarly, \(Z_t \overset{d}{=} Z^{(1)}_{\delta t}\), where \(Z^{(1)}\) is a diffusion process with drift and diffusion coefficients \(c\) and \(g\), respectively.  
We do not require additional ergodicity assumptions on \(Z^{(1)}\) for the slow-scale asymptotics in the limit \(\delta \downarrow 0\).

\subsection{Hamilton-Jacobi-Bellman Equation}

For simplicity and without loss of generality, we take \(\bar\alpha = 0\) throughout.\footnote{The case of nonzero \(\bar\alpha\) can be analyzed analogously, although it leads to more cumbersome expressions that do not provide additional insight into the structure of the optimal trading problem.} In the value function equation, the supremum is taken over admissible strategies that are \(\mathcal{F}_t\)-progressively measurable, square-integrable (i.e., \(\int_0^T u_t^2 \, dt < \infty\) a.s. for all \(T > 0\)), and such that \eqref{eq:x_ou_process}, \eqref{eq:multiscale_model_1}, and \eqref{eq:multiscale_model_2} admit a unique strong solution on \([0, \infty)\).  
The usual dynamic programming principle leads to the HJB equation:
\begin{equation*}
\begin{split}
    \mathcal{L}_2(\sigma(y,z)) v 
    + \left(\frac{1}{\varepsilon} \mathcal{L}_0 v + \frac{1}{\sqrt{\varepsilon}} \mathcal{L}_1 v\right)
    + \sqrt{\delta} \mathcal{M}_1 v 
    + \delta \mathcal{M}_2 v 
    + \sqrt{\frac{\varepsilon}{\delta}} \mathcal{M}_3 v \\
    + \sup_u \left\{ -\lambda q u - \frac{K}{2} u^2 + u v_q + \lambda u v_l \right\} &= 0,
\end{split}
\end{equation*}
where we define the linear operator
\begin{equation*}
    \mathcal{L}_2(\sigma^2(y,z)) = \frac{1}{2} \eta \frac{\partial^2}{\partial x^2} - \kappa x \frac{\partial}{\partial x} - \beta l \frac{\partial}{\partial l} + q x - \beta q l - \frac{\gamma}{2} \sigma^2 q^2,
\end{equation*}
and, following the notation in \citet{fouque2011multiscale},
\begin{align*}
    \mathcal{L}_0 &= \frac{1}{2} a(y)^2 \frac{\partial^2}{\partial y^2} + b(y) \frac{\partial}{\partial y}, \qquad
    \mathcal{L}_1 = \sqrt{\eta} \rho_1 a(y) \frac{\partial^2}{\partial x\partial y},\\
    \mathcal{M}_1 &= \sqrt{\eta} \rho_2 g(z) \frac{\partial^2}{\partial x\partial z}, \quad
    \mathcal{M}_2 = \frac{1}{2} g(z)^2 \frac{\partial^2}{\partial z^2} + c(z) \frac{\partial}{\partial z}, \quad
    \mathcal{M}_3 = \rho_{12} a(y) g(z) \frac{\partial^2}{\partial y\partial z}.
\end{align*}

Note that \(\mathcal{L}_0\) and \(\mathcal{M}_2\) are the infinitesimal generators of the processes \(Y^{(1)}\) and \(Z^{(1)}\), respectively.

We observe that the above maximization problem is quadratic in the control \(u\). Therefore, by substituting the optimal trading rate
\begin{equation*}
    u^* = \frac{1}{K} (\lambda q + v_q + \lambda v_l),
\end{equation*}
we obtain:
\begin{equation}\label{eq:pde_general}
\begin{split}
    \mathcal{L}_2(\sigma(y,z)) v 
    + \frac{1}{2K} (\lambda q + v_q + \lambda v_l)^2 
    + \left(\frac{1}{\varepsilon} \mathcal{L}_0 v + \frac{1}{\sqrt{\varepsilon}} \mathcal{L}_1 v\right) 
    + \sqrt{\delta} \mathcal{M}_1 v 
    + \delta \mathcal{M}_2 v 
    + \sqrt{\frac{\varepsilon}{\delta}} \mathcal{M}_3 v = 0.
\end{split}
\end{equation}

We note that \eqref{eq:pde_general} is a nonlinear partial differential equation, which is not easily solvable either analytically or numerically.  
Our approach is to treat this as a perturbation problem around the special case of \emph{constant volatility}, as studied by \citet{garleanu2013dynamic}.

\subsection{Constant Volatility Solution}\label{subsec: constant}

In the case of constant volatility \(\sigma\), the value function \(v(q,l,x)\) does not depend on the volatility factors \(y\) and \(z\). 
The HJB equation simplifies to
\begin{equation}\label{eq:pde_c}
    \mathcal{L}_2(\sigma)v + \frac{1}{2K} (\lambda q + v_q + \lambda v_l)^2 = 0.
\end{equation}

To solve this, we follow the approach in \citet{garleanu2013dynamic}, assuming a solution of the form:
\begin{align}
    v(q,l,x) = - \frac{1}{2} A_{qq} q^2 + \frac{1}{2} A_{ll} l^2 + \frac{1}{2} A_{xx} x^2 + A_{ql} q l + A_{qx} q x + A_{xl} x l + A_0. \label{eq: ansatz_c} 
\end{align}

Plugging this ansatz (\ref{eq: ansatz_c}) into equation (\ref{eq:pde_c}), we obtain the following algebraic system for the coefficients \(A\)'s:
\begin{align}
    \frac{\rho}{2} A_{qq} - \frac{\gamma}{2} \sigma^2 + \frac{1}{2K} (\lambda - A_{qq} + \lambda A_{ql})^2 &= 0 \label{eq: eq1},\\
    \frac{1}{K} (A_{ql} + \lambda A_{ll})(\lambda - A_{qq} + \lambda A_{ql}) - \beta A_{ql} - \rho A_{ql} - \beta &= 0 \label{eq: eq2},\\
    \frac{1}{K} (A_{qx} + \lambda A_{xl})(\lambda - A_{qq} + \lambda A_{ql}) - \rho A_{qx} - \kappa A_{qx} - 1 &= 0 \label{eq: eq3},\\
    \frac{1}{2K} (A_{ql} + \lambda A_{ll})^2 - \frac{\rho}{2} A_{ll} - \beta A_{ll} &= 0 \label{eq: eq4},\\
    \frac{1}{K} (A_{ql} + \lambda A_{ll})(A_{qx} + \lambda A_{xl}) - \kappa A_{xl} - \beta A_{xl} - \rho A_{xl} &= 0 \label{eq: eq5},\\
    \frac{1}{2K} (A_{qx} + \lambda A_{xl})^2 - \frac{\rho}{2} A_{xx} - \kappa A_{xx} &= 0 \label{eq: eq6},\\
    \frac{\eta}{2} A_{xx} - \rho A_0 &= 0 \label{eq: eq7}.
\end{align}

To compute the coefficients \(A\), we solve this system. It is nonlinear and more complex than in the case without price impact (\(\lambda = \beta = 0\)). In general, there is no closed-form solution. However, for sufficiently large values of the risk-aversion parameter \(\gamma\), the system admits a unique positive solution, as shown by \citet{garleanu2013dynamic}, and verified numerically.\footnote{For every positive value of the parameters \(\rho\), \(\sigma\), \(\lambda\), \(\beta\), \(\eta\), \(\kappa\), and \(K\), we can always choose \(\gamma > 0\) large enough so that the system of equations (\ref{eq: eq1})--(\ref{eq: eq7}) admits four real solutions. Among these, there exists a unique positive solution for \(A_{qq}\).} They primarily analyze the simpler case \(\lambda = \beta = 0\), for which there is a closed-form solution for the coefficients \(A_{qq}, A_{xx}, A_{qx}, A_0\), with the others being zero.

To proceed, we first focus on the subsystem given by equations (\ref{eq: eq1}), (\ref{eq: eq2}), and (\ref{eq: eq4}), which involves the variables \(A_{qq}\), \(A_{ql}\), and \(A_{ll}\). Solving equations (\ref{eq: eq1}) and (\ref{eq: eq4}) for \(A_{qq}\) and \(A_{ll}\), respectively, and substituting into (\ref{eq: eq2}) yields a quartic polynomial in \(A_{ql}\). The roots of this polynomial provide candidate values for \(A_{ql}\), and each root gives corresponding values for \(A_{qq}\) and \(A_{ll}\). Once these three coefficients are determined, the remaining equations become linear and can be solved sequentially.

\subsubsection*{Optimal Trading Strategy}

Solving the system above provides the value function in the form of (\ref{eq: ansatz_c}). The corresponding optimal trading rate is:
\begin{align*}
    \begin{split}
        u^*(q,x,l) &= \frac{1}{K} (\lambda q + v_q + \lambda v_l) \\
        &= \frac{1}{K} \left[(\lambda - A_{qq} + \lambda A_{ql}) q + (A_{ql} + \lambda A_{ll}) l + (A_{qx} + \lambda A_{xl}) x\right].
    \end{split}
\end{align*}

This control can be rewritten in a more interpretable form using the notion of a target portfolio and tracking speed, as introduced by \citet{garleanu2013dynamic, garleanu2016}:
\begin{align*}
    u^* = r^c(\sigma^2) (aim_t^c(\sigma^2) - q),
\end{align*}
where
\begin{align}\label{eq: aim_c}
    r^c(\sigma^2) = \frac{1}{K} (A_{qq} - \lambda - \lambda A_{ql}), \quad \text{and} \quad aim_t^c(\sigma^2) = \frac{(A_{ql} + \lambda A_{ll}) l_t + (A_{qx} + \lambda A_{xl}) x_t}{A_{qq} - \lambda - \lambda A_{ql}},
\end{align}
provided that \(r^c \neq 0\). This formulation reflects that the trader is tracking a time-varying target portfolio \(aim_t^c\) at an adjustment speed \(r^c\). We emphasize the dependence of \(r^c\) and \(aim_t^c\) on the volatility parameter \(\sigma\), as this dependence will be extended in the stochastic volatility case.

\subsection{Small Price Impact Approximation}\label{sec:approx}

The system (\ref{eq: eq1}) - (\ref{eq: eq7}) we must solve to find the leading order term is nonlinear. Therefore, in this subsection, we propose an analytical technique to simplify and accelerate computations. Empirical evidence from \citet{webster2023}, as well as the references cited therein, suggests that the price impact is relatively small compared to the size of the position held in the stock. Thus, we introduce the scalling paramter \(\theta\) into the dynamics of the  price impact process as:
\begin{align*}
    dl_t = -\theta \beta l_t\, dt + \theta \lambda\, dq_t,
\end{align*}
where \(\theta\) is assumed to be small.

We introduce the notation \(A^{\theta}\) for the solution to the algebraic system, with the replacements \(\beta \mapsto \theta \beta\) and \(\lambda \mapsto \theta \lambda\).

We then expand each coefficient \(A^\theta\) in a power series in \(\theta\):
\begin{align*}
    A_{qq}^\theta = A_{qq}^{(0)} + \theta A_{qq}^{(1)} + \theta^2 A_{qq}^{(2)} + \cdots.
\end{align*}

Plugging this ansatz and expansion into the system yields a polynomial in \(\theta\). The coefficients of each power of \(\theta\) must vanish identically, allowing us to solve recursively.

\subsubsection{Zeroth Order Term \(\theta^0\)}

Setting \(\theta\) equal to zero, the system matches the result in \citet{garleanu2013dynamic}, which neglects price impact. Hence, we obtain the simpler system:
\begin{align}
\begin{split}
    \frac{1}{2K} {A_{qq}^{(0)}}^2 + \frac{1}{2} \rho A_{qq}^{(0)} - \frac{\gamma \sigma^2}{2} &= 0, \label{eq:1}\\
    1 - \rho A_{qx}^{(0)} - \frac{1}{K} A_{qq}^{(0)} A_{qx}^{(0)} - \kappa A_{qx}^{(0)} &= 0, \\
    \frac{1}{2K} {A_{qx}^{(0)}}^2 - \kappa A_{xx}^{(0)} - \frac{1}{2} \rho A_{xx}^{(0)} &= 0, \\
    \frac{\eta}{2} A_{xx}^{(0)} - \rho A_0^{(0)} &= 0,
\end{split}
\end{align}
with  \(A_{ql}^{(0)} = A_{ll}^{(0)} = A_{xl}^{(0)} = 0\).

In the above system, \eqref{eq:1} is scalar quadratic for \(A_{qq}^{(0)}\), and the other three are linear in the remaining coefficients. Thus, the solution is given by
\begin{align*}
    A_{qq}^{(0)} = \frac{K}{2} \left( \sqrt{\rho^2 + \frac{4 \gamma \sigma^2}{K}} - \rho \right), \quad A_{qx}^{(0)} = \left( \kappa + \rho + \frac{A_{qq}^{(0)}}{K} \right)^{-1}, \quad  A_{xx}^{(0)} = \frac{{A_{qx}^{(0)}}^2}{K (2\kappa + \rho)}, \quad A_0^{(0)} = \frac{\eta}{2\rho} A_{xx}^{(0)}.
\end{align*}

\subsubsection{First Order Term \(\theta^1\)}

We now consider the \(\theta^1\) term. The corresponding system of equations is:
\begin{align*}
    \frac{\rho}{2} A_{qq}^{(1)} - \frac{1}{K} \left( \lambda - A_{qq}^{(1)} + \lambda A_{ql}^{(0)} \right) A_{qq}^{(0)} &= 0, \\
    \frac{1}{K} \left( \lambda - A_{qq}^{(1)} + \lambda A_{ql}^{(0)} \right) A_{ql}^{(0)} - \beta A_{ql}^{(0)} - \rho A_{ql}^{(1)} - \beta - \frac{1}{K} \left( A_{ql}^{(1)} + \lambda A_{ll}^{(0)} \right) A_{qq}^{(0)} &= 0, \\
    \frac{1}{K} \left( \lambda - A_{qq}^{(1)} + \lambda A_{ql}^{(0)} \right) A_{qx}^{(0)} - \rho A_{qx}^{(1)} - \kappa A_{qx}^{(1)} - \frac{1}{K} \left( A_{qx}^{(1)} + \lambda A_{xl}^{(0)} \right) A_{qq}^{(0)} &= 0, \\
    \frac{1}{K} \left( A_{ql}^{(1)} + \lambda A_{ll}^{(0)} \right) A_{ql}^{(0)} - \frac{\rho}{2} A_{ll}^{(1)} - \beta A_{ll}^{(0)} &= 0, \\
    \frac{1}{K} \left( A_{ql}^{(1)} + \lambda A_{ll}^{(0)} \right) A_{qx}^{(0)} - \kappa A_{xl}^{(1)} - \rho A_{xl}^{(1)} - \beta A_{xl}^{(0)} + \frac{1}{K} \left( A_{qx}^{(1)} + \lambda A_{xl}^{(0)} \right) A_{ql}^{(0)} &= 0, \\
    \frac{1}{K} \left( A_{qx}^{(1)} + \lambda A_{xl}^{(0)} \right) A_{qx}^{(0)} - \frac{\rho}{2} A_{xx}^{(1)} - \kappa A_{xx}^{(1)} &= 0,\\
    \frac{\eta}{2} A_{xx}^{(1)} - \rho A_{0}^{(1)} &= 0,
\end{align*}
which is linear in the unknown \(A^{(1)}\) terms. Solving we find:
\begin{align*}
    A_{qq}^{(1)} = \frac{2 \lambda A_{qq}^{(0)}}{2 A_{qq}^{(0)} + K \rho}, \qquad A_{ql}^{(1)} = -\frac{\beta K}{\rho K + A_{qq}^{(0)}}, \qquad A_{qx}^{(1)} = {A_{qx}^{(0)}}^2 \left( \lambda - A_{qq}^{(1)} \right)\\
    A_{ll}^{(1)} = 0, \qquad A_{xl}^{(1)} = \frac{A_{qx}^{(0)} A_{ql}^{(1)}}{K(\kappa+\rho)}, \qquad A_{xx}^{(1)} = \frac{2 A_{qx}^{(0)} A_{qx}^{(1)}}{K\left(2\kappa +\rho \right)}, \qquad A_{0}^{(1)} = \frac{\eta}{2\rho} A_{xx}^{(1)}.
\end{align*}

The second-order term is provided in the appendix. The accuracy of the first order approximation is numerically demonstrated in Section \ref{sec:numeric}.

\section{Single Fast Factor}\label{sec:fast}

We first analyze the optimal trading problem when there is only one fast volatility factor, \(\sigma = \sigma(Y_t)\). In this case, the PDE of the value function \(v(q,l,x,y)\) is:
\begin{equation}\label{eq:pde_fast}
\begin{split}
    \mathcal{L}_2(\sigma^2(y)) v^\varepsilon + \frac{1}{2K} (\lambda q + v^\varepsilon_q + \lambda v^\varepsilon_l)^2 +\left(\frac{1}{\varepsilon} \mathcal{L}_0 v^\varepsilon + \frac{1}{\sqrt{\varepsilon}}\mathcal{L}_1 v^\varepsilon\right) = 0.
\end{split}
\end{equation}

This is a singular perturbation problem in the limit \(\varepsilon \downarrow 0\), and we approach it by constructing an asymptotic expansion for the value function:
\begin{align*}
    v^\varepsilon(q,l,x,y) = v^{(0)}(q,l,x) + \sqrt{\varepsilon}\, v^{(1)}(q,l,x) + \varepsilon\, v^{(2)}(q,l,x,y) + \varepsilon^{3/2} v^{(3)}(q,l,x,y) + \cdots
\end{align*}

By substituting this expansion into \eqref{eq:pde_fast} and collecting terms by powers of \(\varepsilon\), we analyze the system order-by-order in the subsections below.

\subsection{Zeroth Order Term \(v^{(0)}\)}

To begin with, the \(\varepsilon^{-1}\) order yields the PDE \(\mathcal{L}_0 v^{(0)} = 0\). Since \(\mathcal{L}_0\) contains derivatives only with respect to \(y\), we choose \(v^{(0)}\) to be independent of \(y\); hence, \(v^{(0)}_y \equiv 0\).

Next, the \(\varepsilon^{-1/2}\) order gives:
\begin{align*}
    \mathcal{L}_0 v^{(1)} + \mathcal{L}_1 v^{(0)} = 0.
\end{align*}
Since  \(\mathcal{L}_1 v^{(0)}=0\), we again choose \(v^{(1)}\) to be independent of \(y\).

Now consider the \(\varepsilon^0\) order:
\begin{align}
    \mathcal{L}_0 v^{(2)} +\mathcal{L}_1 v^{(1)} + \ltwo(\sigma(y)) v^{(0)} + \frac{1}{2K} (\lambda q + v_q^{(0)} + \lambda v_l^{(0)})^2= 0,\label{eq:pde_eps0}
\end{align}
and observe that \(\mathcal{L}_1 v^{(1)} = 0.\)

We view \eqref{eq:pde_eps0} as a Poisson equation for \(v^{(2)}\). The Fredholm alternative (solvability condition) requires:
\begin{align*}
    \Big\langle \ltwo(\sigma^2(y))v^{(0)} + \frac{1}{2K} (\lambda q + v_q^{(0)} + \lambda v_l^{(0)})^2\Big\rangle = 0,
\end{align*}
where \(\langle \cdot \rangle\) denotes expectation under the invariant distribution \(\Phi\) of the ergodic process \(Y_t^{(1)}\):
\begin{align*}
    \langle h \rangle = \int h(y)\, \Phi(dy).
\end{align*}

Using the fact that \(v^{(0)}\) is independent of \(y\), we obtain the simplified solvability condition:
\begin{equation}\label{eq:pde_c_2}
    \ltwo(\langle\sigma^2\rangle)v^{(0)} + \frac{1}{2K} (\lambda q + v^{(0)}_q + \lambda v^{(0)}_l)^2 = 0.
\end{equation}
This is identical to the PDE from the constant-volatility case (Section~\ref{subsec: constant}), except with \(\sigma\) replaced by \(\langle \sigma \rangle\). Therefore, \(v^{(0)}\) can be assumed to have the same quadratic form:
\begin{align*}
    v^{(0)}(q,l,x) &= -\frac{1}{2} A_{qq} q^2 + \frac{1}{2} A_{ll} l^2 + \frac{1}{2} A_{xx} x^2 + A_{ql} ql + A_{qx} qx + A_{xl} xl + A_0.
\end{align*}§

We then return to \eqref{eq:pde_eps0} and subtracting the left hand side of \eqref{eq:pde_c_2}, we obtain:
\begin{align}\label{eq:v2_p}
    \mathcal{L}_0 v^{(2)} = \frac{\gamma}{2} \left( \sigma^2(y) - \langle \sigma^2 \rangle \right) q^2 =: h(y).
\end{align}

This Poisson equation has the general solution:
\begin{align*}
    v^{(2)}(q,y) = -\int_0^\infty \mathbf{P}_t h(y)\, dt + \mathcal{C}(q,l,x),
\end{align*}
where \(\mathcal{C}\) is independent of \(y\), and \(\mathbf{P}_t h(y) := \mathbb{E}[h(Y_t^{(1)}) \mid Y_0^{(1)} = y]\). See \citet[Section 3.2]{fouque2011multiscale}.

\subsection{First Order Term \(v^{(1)}\)}

The \(\sqrt{\varepsilon}\) order in the expansion yields:
\begin{equation*}
    -\rho v^{(1)} - \beta l v_l^{(1)} + \frac{1}{K} \left(v_q^{(1)} + \lambda v_l^{(1)}\right) \left(\lambda q + v_q^{(0)} + \lambda v_l^{(0)}\right)- \kappa x v_x^{(1)} + \frac{1}{2} \eta v_{xx}^{(1)} + \mathcal{L}_0 v^{(3)} + \mathcal{L}_1 v^{(2)}=0.
\end{equation*}

This is a Poisson equation for \(v^{(3)}\). Its solvability condition is:
\begin{equation*}
    -\rho v^{(1)} - \beta l v_l^{(1)} + \frac{1}{K} \left(v_q^{(1)} + \lambda v_l^{(1)}\right) \left(\lambda q + v_q^{(0)} + \lambda v_l^{(0)}\right) - \kappa x v_x^{(1)} + \frac{1}{2} \eta v_{xx}^{(1)} = 0.
\end{equation*}

This is a linear and homogeneous PDE for \(v^{(1)}\), which admits the trivial solution \(v^{(1)} \equiv 0\). As a result, we also have \(\mathcal{L}_0 v^{(3)} = 0\) and we choose \(v^{(3)}\) to be independent of \(y\).

\subsection{Second Order Term \(v^{(2)}\)} \label{subsec:examples}

At order \(\varepsilon\), the expansion gives:
\begin{align*}
    &-\rho v^{(2)} - \beta l v_l^{(2)} 
    + \frac{1}{2K} \left(v_q^{(1)} + \lambda v_l^{(1)}\right)^2 
    + \frac{1}{K} \left(v_q^{(2)} + \lambda v_l^{(2)}\right) \left(\lambda q + v_q^{(0)} + \lambda v_l^{(0)}\right) \\
    &\quad - \kappa x v_x^{(1)} + \frac{1}{2} \eta v_{xx}^{(1)}
    + \mathcal{L}_0 v^{(4)} + \mathcal{L}_1 v^{(3)} = 0.
\end{align*}

Since \(v^{(1)} = 0\) and \(v^{(3)}_y = 0\), this reduces to a linear homogeneous equation for \(\mathcal{C}(q,l,x)\):
\begin{align*}
    -\rho \mathcal{C} - \beta l \mathcal{C}_l + \frac{1}{K} \left(\mathcal{C}_q + \lambda \mathcal{C}_l\right) \left(\lambda q + v_q^{(0)} + \lambda v_l^{(0)}\right) = 0.
\end{align*}

We may choose \(\mathcal{C} \equiv 0\) without loss of generality. Therefore, the leading-order correction to the value function is:
\begin{align}\label{eq:fast_v2_final}
    v^{(2)}(q,y) = -\frac{1}{2} q^2 \phi(y), \quad \text{where we define }\phi(y) := \gamma \int_0^\infty \mathbf{P}_t\left(\sigma^2(y) - \langle \sigma^2 \rangle\right) dt.
\end{align}

\begin{example}\label{ex:ou_fast_vol}
Suppose \(\sigma^2(y) = y\), and the volatility factor \(Y^{(1)}_t\) follows the Cox-Ingersoll-Ross (CIR) process \citeyearpar{cox1985theory}, given by:
\begin{equation*}
    b(y) = \theta(\mu - y), \quad a(y) = \psi \sqrt{y}.
\end{equation*}
Applying the transition semigroup \(\mathbf{P}_t\) to the function \(h(y) = \frac{\gamma}{2} (y - \mu) q^2\), we compute:
\[
\begin{split}
\mathbf{P}_t h(y) &= \mathbb{E}\left[h(Y^{(1)}_t)\, \big|\, Y^{(1)}_0 = y\right] \\
&= \frac{\gamma}{2} q^2\, \mathbb{E}\left[Y^{(1)}_t - \mu \,\big|\, Y^{(1)}_0 = y\right] \\
&= \frac{\gamma}{2} q^2\, e^{-\theta t}(y - \mu).
\end{split}
\]
Then by integrating as in equation~\eqref{eq:fast_v2_final}, we find the second-order correction:
\[
v^{(2)} = -\frac{\gamma}{2\theta} (y - \mu) q^2.
\]
One can directly verify that this expression solves the Poisson equation~\eqref{eq:v2_p}.
\end{example}

\begin{example}\label{ex:expOU_fast_vol}
As an alternative, consider the exponential Ornstein-Uhlenbeck (OU) stochastic volatility model studied by \citet{masoliver2006multiple}. Here, \(\sigma(y) = m e^y\), and the volatility factor \(Y^{(1)}_t\) follows:
\[
b(y) = -\theta y, \quad a(y) = \widehat\sigma.
\]
We apply the transition semigroup \(\mathbf{P}_t\) to:
\[
h(y) = \frac{\gamma}{2} q^2 \left(\sigma^2(y) - \langle \sigma^2 \rangle \right) = \frac{\gamma}{2} q^2 \left(m^2 e^{2y} - \langle m^2 e^{2y} \rangle \right).
\]
Using known properties of the OU process, we compute:
\[
\begin{split}
\mathbf{P}_t h(y) &= \mathbb{E}\left[h(Y^{(1)}_t)\, \big|\, Y^{(1)}_0 = y \right] \\
&= \frac{\gamma}{2} q^2 m^2 e^{k^2/\alpha} \left( e^{2y e^{-\alpha t} - \frac{k^2}{\alpha} e^{-2\alpha t}} - 1 \right).
\end{split}
\]
While the integral \(\int_0^\infty \mathbf{P}_t h(y)\, dt\) does not admit a simple closed-form, an approximation can be derived for small fluctuations around the mean:
\[
\begin{split}
v^{(2)} &= - \frac{\gamma}{2} q^2 m^2 e^{k^2/\alpha} \int_0^\infty \left(e^{2y e^{-\alpha t} - \frac{k^2}{\alpha} e^{-2\alpha t}} - 1\right)\, dt \\
&\approx - \frac{\gamma}{4\alpha} q^2 m^2 e^{k^2/\alpha} \left( 4y\, \Gamma\left(\tfrac{1}{2},\tfrac{k}{\alpha^2}\right) - \left[\widehat\gamma + \Gamma\left(0,\tfrac{k^2}{\alpha}\right) + \log\left(\tfrac{k^2}{\alpha}\right)\right] \right),
\end{split}
\]
where \(\widehat\gamma \approx 0.5772\) is the Euler–Mascheroni constant, and \(\Gamma(a, z)\) is the upper incomplete gamma function:
\[
\Gamma(a, z) = \int_z^\infty t^{a-1} e^{-t} dt.
\]
\end{example}

\subsection{Optimal Trading Strategy}

Incorporating the correction term into the control, we obtain:
\begin{align*}
\begin{split}
    u^*(q,l,x,y) &= \frac{1}{K} \left[(\lambda - A_{qq} + \lambda A_{ql}) q + (A_{ql} + \lambda A_{ll}) l + (A_{qx} + \lambda A_{xl}) x\right] - \frac{1}{K} \varepsilon q \phi(y) \\
    &= \left(r^c(\langle \sigma^2 \rangle) + \frac{1}{K} \varepsilon \phi(y)\right)\left(\frac{(A_{ql} + \lambda A_{ll}) l + (A_{qx} + \lambda A_{xl}) x}{A_{qq} - \lambda - \lambda A_{ql} + \varepsilon\phi(y)} - q\right),
\end{split}
\end{align*}
where, we write \(u^*_t := r^f (aim_t^f - q)\) with:
\begin{align*}
    r^f := r^c(\langle \sigma^2 \rangle) + \frac{1}{K} \varepsilon \phi(Y_t),
\end{align*}
and
\begin{align*}
    aim_t^f &:= \frac{(A_{ql} + \lambda A_{ll}) l_t + (A_{qx} + \lambda A_{xl}) x_t}{A_{qq} - \lambda - \lambda A_{ql} + \varepsilon\phi(Y_t)} =\left(1 - \varepsilon \frac{\phi(Y_t)}{A_{qq} - \lambda - \lambda A_{ql}} + \cdots \right)\times aim_t^c(\langle \sigma^2 \rangle),
\end{align*}
and recall that \(r^c\) and \(aim^c\) were given in \eqref{eq: aim_c}, assuming again that \(r^c \neq 0\).

Thus, the principal correction to the optimal trading rate is given by:
\begin{equation*}
    u^{(2)}(q,y) = -\frac{q}{K} \phi(y).
\end{equation*}

\noindent\textbf{Example~\ref{ex:ou_fast_vol} (continued).} In the case where the volatility factor is governed by a Cox-Ingersoll-Ross process, this correction simplifies to:
\begin{equation*}
    u^{(2)}(q,y) = - \frac{\gamma}{\theta K} q (y - \mu).
\end{equation*}
This implies that the investor should optimally reduce exposure when the current volatility level is above its long-term average.

In the case of a fast CIR volatility factor, the corrected strategy parameters are:
\[
\begin{split}
r^f &= r^c + \varepsilon \frac{\gamma}{\theta K} (y - \mu), \\
aim_t^f &= aim_t^c \left(1 - \varepsilon \frac{\gamma}{\theta (A_{qq} - \lambda - \lambda A_{ql})} (y - \mu)\right).
\end{split}
\]

In words, the optimal trading speed \(r^f\) increases with the current volatility level \(y\), while the size of the target portfolio \(aim_t^f\) is scaled down when volatility is above its long-run mean. These effects are mitigated by a larger transaction cost \(K\) or a lower risk-aversion coefficient \(\gamma\).

\section{Single Slow Factor}\label{sec:slow}

In this section, we consider the case where only one slow-scale stochastic volatility factor is present in the model. In this case, the PDE of the value function \(v(q,l,x,z)\) is:
\begin{equation}\label{eq:pde_slow}
    \mathcal{L}_2(\sigma(z)) v^\delta + \frac{1}{2K} (\lambda q + v_q^{\delta} + \lambda v_l^{\delta})^2 + \delta \mathcal{M}_2 v^{\delta} + \sqrt{\delta} \mathcal{M}_1 v^{\delta} = 0.
\end{equation}
This is structurally similar to equation~\eqref{eq:pde_general} but without the \(y\)-dependence. To solve the PDE, we apply the following asymptotic expansion:
\begin{align*}
    v^{\delta}(q,l,x,z) = v^{(0)}(q,l,x,z) + \sqrt{\delta} v^{(1)}(q,l,x,z) + \delta v^{(2)}(q,l,x,z) + \cdots.
\end{align*}

\subsection{Zeroth Order Term \(\delta^0\)}

Substituting the expansion into the PDE equation and collecting terms of like powers of \(\delta\), we start with the constant term (i.e., setting \(\delta = 0\)) to obtain:
\begin{equation*}
    \mathcal{L}_2(\sigma^2(z)) v + \frac{1}{2K} (\lambda q + v_q + \lambda v_l)^2 = 0.
\end{equation*}

This PDE resembles equation~\eqref{eq:pde_c}, with the only difference being the explicit \(z\)-dependence of \(\sigma\). Therefore, the problem reduces to a parameterized version of the constant volatility case. From section \ref{subsec: constant}, we have the solution:
\begin{align*}
    v^{(0)}(q,l,x) = &-\frac{1}{2} A_{qq}(z) q^2 + \frac{1}{2} A_{ll}(z) l^2 + \frac{1}{2} A_{xx}(z) x^2 + A_{ql}(z) ql + A_{qx}(z) qx + A_{xl}(z) xl + A_0(z),
\end{align*}
where we show the dependence of coefficients on \(z\) inherited from \(\sigma(z)\).

\subsection{First Order Term \(\sqrt{\delta}\)}

At order \(\sqrt{\delta}\), the resulting PDE is:
\begin{align*}
    -\rho v^{(1)} - \beta l v_l^{(1)} - \kappa x v_x^{(1)} + \frac{1}{2} \eta v_{xx}^{(1)} + \frac{1}{K} (v_q^{(1)} + \lambda v_l^{(1)})(\lambda q + v_q^{(0)} + \lambda v_l^{(0)}) + \mathcal{M}_1 v^{(0)} = 0.
\end{align*}
We assume a linear form for \(v^{(1)}\):
\begin{align*}
    v^{(1)}(q,l,x,z) = B_q(z) q + B_l(z) l + B_x(z) x.
\end{align*}
By plugging this linear ansatz into the previous equation, we obtain a polynomial in the variables \(q\), \(l\), and \(x\). Since this polynomial is identically zero, equating the coefficients of each monomial to zero yields the following system:
\begin{align*}
\frac{1}{K} (B_q(z) + \lambda B_l(z))(\lambda - A_{qq}(z) + \lambda A_{ql}(z)) 
- \rho B_q(z) + \sqrt{\eta} \rho_2 A_{qx}^{\prime}(z) g(z) &= 0, \\
\frac{1}{K} (A_{ql}(z) + \lambda A_{ll}(z))(B_q(z) + \lambda B_l(z)) 
- (\rho + \beta) B_l(z) + \sqrt{\eta} \rho_2 A_{xl}^{\prime}(z) g(z) &= 0, \\
\frac{1}{K} (A_{qx}(z) + \lambda A_{xl}(z))(B_q(z) + \lambda B_l(z)) 
- (\rho + \kappa) B_x(z) + \sqrt{\eta} \rho_2 A_{xx}^{\prime}(z) g(z) &= 0.
\end{align*}
This is a linear system in the variables \(B_q(z)\), \(B_l(z)\), and \(B_x(z)\). We define
\begin{align*}
\Gamma(z) = \beta \lambda - \rho A_{qq}(z) - \beta A_{qq}(z) + \lambda \rho - K \rho^2 + \beta \lambda A_{ql}(z) + 2 \lambda \rho A_{ql}(z) - K \beta \rho + \lambda^2 \rho A_{ll}(z).
\end{align*}
In the case when \(\lambda = \beta = 0\), we can easily see that \(\Gamma(z) > 0\) for all \(\sigma(z) > 0\). Hereafter, we assume that price impact is small enough for this to remain the case.
The solution to the linear system is then given by
\begin{align} \label{eq: coef_B}
\begin{split}
B_q(z) = -\sqrt{\eta} \rho_2 g(z) \Gamma(z)^{-1} \Big(&
A_{xl}^{\prime}(z) \lambda^2 
- A_{ql}(z) A_{qx}^{\prime}(z) \lambda 
- A_{qq}(z) A_{xl}^{\prime}(z) \lambda
+ A_{qx}^{\prime}(z) K \beta \\
&+ A_{qx}^{\prime}(z) K \rho 
- A_{ll}(z) A_{qx}^{\prime}(z) \lambda^2 + A_{ql}(z) A_{xl}^{\prime}(z) \lambda^2
\Big), \\
B_l(z) = -\sqrt{\eta} \rho_2 g(z) \Gamma(z)^{-1} \Big(&
A_{ql}(z) A_{qx}^{\prime}(z) 
- A_{xl}^{\prime}(z) \lambda 
+ A_{qq}(z) A_{xl}^{\prime}(z)
+ A_{ll}(z) A_{qx}^{\prime}(z) \lambda \\
&- A_{ql}(z) A_{xl}^{\prime}(z) \lambda 
+ A_{xl}^{\prime}(z) K \rho
\Big), \\
B_x(z) = -\sqrt{\eta} \frac{\rho_2}{\kappa + \rho} g(z) \Gamma(z)^{-1} \Big(&
A_{qx}(z) A_{qx}^{\prime}(z) \beta 
+ A_{qq}(z) A_{xx}^{\prime}(z) \beta
+ A_{qx}(z) A_{qx}^{\prime}(z) \rho
+ A_{qq}(z) A_{xx}^{\prime}(z) \rho \\
&- A_{xx}^{\prime}(z) \beta \lambda 
- A_{xx}^{\prime}(z) \lambda \rho 
+ A_{xx}^{\prime}(z) K \rho^2
- A_{ll}(z) A_{xx}^{\prime}(z) \lambda^2 \rho\\ 
&+ A_{xl}(z) A_{xl}^{\prime}(z) \lambda^2 \rho
- A_{ql}(z) A_{xx}^{\prime}(z) \beta \lambda 
+ A_{qx}^{\prime}(z) A_{xl}(z) \beta \lambda \\
&- 2 A_{ql}(z) A_{xx}^{\prime}(z) \lambda \rho 
+ A_{qx}(z) A_{xl}^{\prime}(z) \lambda \rho
+ A_{qx}^{\prime}(z) A_{xl}(z) \lambda \rho \\
&+ A_{xx}^{\prime}(z) K \beta \rho
\Big).
\end{split}
\end{align}

\subsection{Optimal trading strategy}

Under these model settings, should we take into consideration only the first correction term, the optimal trading strategy is
\begin{align*}
\begin{split}
u^* = \frac{1}{K} \Big((\lambda - A_{qq}(z) + \lambda A_{ql}(z)) q + (A_{ql}(z) + \lambda A_{ll}(z)) l + (A_{qx}(z) + \lambda A_{xl}(z)) x\Big) + \frac{\sqrt{\delta}}{K} (B_q(z) + \lambda B_l(z))
\end{split}
\end{align*}

where \(u^* := u^*(q,l,x,z)\).

Using the same representation as in the previous cases with constant and slow-scale volatility, we can write the optimal control as
\begin{align*}
\begin{split}
u^* = \frac{1}{K} \Big((\lambda - A_{qq}(z) + \lambda A_{ql}(z)) q + (A_{ql}(z) + \lambda A_{ll}(z)) l + (A_{qx}(z) + \lambda A_{xl}(z)) x\Big) + \frac{\sqrt{\delta}}{K} (B_q(z) + \lambda B_l(z)).
\end{split}
\end{align*}
As before, we write  \(u^*_t = r^s(Z_t)(aim_t^s - q)\), with
\begin{align*}
r^s := r^c(\sigma^2(Z_t)) \quad \text{and} \quad aim_t^s := aim_t^c(\sigma^2(Z_t)) + \sqrt{\delta} \frac{B_q(Z_t) + \lambda B_l(Z_t)}{A_{qq} - \lambda - \lambda A_{ql}}.
\end{align*}
When price impact is relatively small, it follows from a straight-forward calculation that $B_q$ and $\rho_2$ have opposite signs, provided that the function $\sigma$ is monotonically increasing in the volatility factor $Z_t$.
When the correlation $\rho_2$ between the volatility and return factors is positive, the investor optimally decreases his trading rate as he anticipates a higher return estimate is accompanied by a higher volatility.
On the other hand, if the return-volatility correlation $\rho_2$ is negative, the investor optimally increases his trading rate $u$.

In Appendix~\ref{ap:derivatives}, we provide a straightforward method for computing the necessary derivatives required to evaluate the \(B\) terms in formulas~\eqref{eq: coef_B}. In Appendix~\ref{ap: fast_2}, we present explicit formulas for the second-order approximation in the case of a single slow factor.

\section{Multiscale multifactor volatility model}\label{sec:multi}

We consider the enriched model with both fast and slow volatility factors. In our analysis below, we follow the same reasoning as in the previous sections and derive similar results with a few modifications. The derived PDE is \eqref{eq:pde_general} plunging \(v^{\varepsilon,\delta}\) instead of \(v\).

To deal with this PDE, we introduce the asymptotic expansion with respect to both \(\delta\) and \(\varepsilon\). We begin with the expansion in the slow scale factor:
\begin{align*}
    v^{\varepsilon, \delta} = v^{\varepsilon, 0} + \sqrt{\delta} \, v^{\varepsilon, 1} + \delta \, v^{\varepsilon, 2} + \cdots.
\end{align*}

Setting \(\delta = 0\), we obtain the following PDE:
\begin{equation*}
    \ltwo(\sigma^2(y,z))v^{\varepsilon,0} + \frac{1}{2K} \left( \lambda q + v_q^{\varepsilon,0} + \lambda v_l^{\varepsilon,0} \right)^2 + \frac{1}{\varepsilon} \mathcal{L}_0 v^{\varepsilon,0} + \frac{1}{\sqrt{\varepsilon}} \mathcal{L}_1 v^{\varepsilon,0}.
\end{equation*}

Next, we expand with respect to the fast volatility factor:
\begin{align*}
    v^{\varepsilon, 0} = v^{(0)} + \sqrt{\varepsilon} \, v^{(1,0)} + \varepsilon \, v^{(2,0)} + \cdots.
\end{align*}

This is the same setting as in Section~\ref{sec:fast}. Hence, using the same analysis, we obtain that \(v^{(0)}\) is the solution from the constant volatility case, where \(\Bar{\sigma}^2(z) := \langle \sigma^2(\cdot, z) \rangle\).

The first and second correction terms, as derived in Section~\ref{sec:fast}, are:
\begin{align*}
    v^{(1,0)} &= 0, \\
    v^{(2,0)}(q,z,y) &= -\frac{\gamma}{2} q^2 \int_0^{+\infty} \mathbf{P}_t \left( \sigma^2(y,z) - \Bar{\sigma}^2(z) \right) dt := -\frac{1}{2} q^2 \phi(y,z).
\end{align*}

We now consider the first-order correction in the slow scale expansion, at order \(\sqrt{\delta}\). This yields the PDE:
\begin{align}
\begin{split}
    0 &= \frac{1}{K} \left( v_q^{\varepsilon,1} + \lambda v_l^{\varepsilon,1} \right) \left( \lambda q + v_q^{\varepsilon,0} + \lambda v_l^{\varepsilon,0} \right) - \rho v^{\varepsilon,1} - \beta l v_l^{\varepsilon,1} - \kappa x v_x^{\varepsilon,1} + \frac{1}{2} v_{xx}^{\varepsilon,1} \\
    &\quad + \frac{1}{\varepsilon} \mathcal{L}_0 v^{\varepsilon,1} + \frac{1}{\sqrt{\varepsilon}} \sqrt{\eta} \rho_1 a(y) v_{xy}^{\varepsilon,1} + \sqrt{\eta} \rho_2 g(z) v_{xz}^{\varepsilon,0} + \sqrt{\frac{\delta}{\varepsilon}} \rho_{12} a(y) g(z) v_{yz}^{\varepsilon,0}. \label{eq:mulit_1_0}
\end{split}
\end{align}

Expanding further:
\begin{align*}
    v^{\varepsilon,1} = v^{(0,1)} + \sqrt{\varepsilon} \, v^{(1,1)} + \varepsilon \, v^{(2,1)} + \cdots.
\end{align*}

Considering the \(\varepsilon^{-1}\) order in equation~\eqref{eq:mulit_1_0}, we obtain \(\mathcal{L}_0 v^{(0,1)} = 0\), implying that \(v^{(0,1)} = v^{(0,1)}(q,l,x,z)\), independent of \(y\). Similarly, from the \(\varepsilon^{-1/2}\) order, \(\mathcal{L}_0 v^{(1,1)} = 0\), so \(v^{(1,1)} = v^{(1,1)}(q,l,x,z)\).

Now, collecting the constant order terms, we derive the following PDE:
\begin{align*}
\begin{split}
    0 &= \frac{1}{K} \left( v_q^{(0,1)} + \lambda v_l^{(0,1)} \right) \left( \lambda q + v_q^{(0)} + \lambda v_l^{(0)} \right) - \rho v^{(0,1)} - \beta l v_l^{(0,1)} - \kappa x v_x^{(0,1)} + \frac{1}{2} v_{xx}^{(0,1)} \\
    &\quad + \mathcal{L}_0 v^{(2,1)} + \mathcal{L}_1 v^{(1,1)} + \mathcal{M}_1 v^{(0)} + \mathcal{M}_3 v^{(1,0)}.
\end{split}
\end{align*}

Recognizing this as a Poisson equation in \(v^{(2,1)}\), we apply the solvability condition:
\begin{align*}
    0 = \frac{1}{K} \left( v_q^{(0,1)} + \lambda v_l^{(0,1)} \right) \left( \lambda q + v_q^{(0)} + \lambda v_l^{(0)} \right) - \rho v^{(0,1)} - \beta l v_l^{(0,1)} - \kappa x v_x^{(0,1)} + \frac{1}{2} v_{xx}^{(0,1)} + \mathcal{M}_1 v^{(0)}.
\end{align*}

This matches the PDE equation for the fast scale case. Therefore, by the same reasoning, we conclude:
\begin{align*}
    v^{(0,1)}(q,l,x,z) = B_q(z) \, q + B_l(z) \, l + B_x(z) \, x,
\end{align*}
where the coefficients \(B_q(z)\), \(B_l(z)\), and \(B_x(z)\) are given in formulas~(\ref{eq: coef_B}).

Here, \(A_{qq}(z)\), \(A_{ll}(z)\), \(A_{xx}(z)\), \(A_{ql}(z)\), \(A_{qx}(z)\), and \(A_{xl}(z)\) are defined as in the constant volatility case with volatility \(\Bar{\sigma}^2(z)\).

\subsection*{Optimal Trading Strategy}

From the previous analysis, we conclude that the optimal trading rate, incorporating corrections to the constant volatility case up to orders \(\varepsilon\) and \(\sqrt{\delta}\), is given by:
\begin{align*}
\begin{split}
    {u^{\varepsilon,\delta}}^* =\;& \frac{1}{K} \left[ ( \lambda - A_{qq}(z) + \lambda A_{ql}(z) ) q + (A_{ql}(z) + \lambda A_{ll}(z)) l + (A_{qx}(z) + \lambda A_{xl}(z)) x \right] \\
    &+ \frac{1}{K} \sqrt{\delta} \left( B_q(z) + \lambda B_l(z) \right) - \frac{1}{K} \varepsilon q \, \phi(y,z),
\end{split}
\end{align*}
where \(A_{qq}(z), A_{ql}(z), A_{ll}(z), A_{qx}(z), A_{xl}(z)\) are derived from the solution of the system in Section~\ref{subsec: constant}, and \(B_q(z), B_l(z)\) are as derived in Section~\ref{sec:slow}.

Using the same notation introduced in earlier Sections, we express the optimal strategy as:
\begin{align*}
    u^* =\; r^m\left( aim_t^m - q \right),
\end{align*}
where we define: 
\[ r^m := r^c(\Bar{\sigma}^2(Z_t)) + \frac{1}{K} \varepsilon \phi(Y_t, Z_t) \quad \text{and}\]
\begin{align*}
    \quad aim_t^m := aim_t^c(\Bar{\sigma}^2(Z_t)) \times \left(1 - \epsilon \frac{\phi(Y_t)}{A_{qq} - \lambda - \lambda A_{ql}} \right)  + \sqrt{\delta} \frac{B_q(Z_t) + \lambda B_l(Z_t)}{A_{qq}(Z_t) - \lambda - \lambda A_{ql}(Z_t) + \varepsilon \phi(Y_t)}.
\end{align*}

\begin{remark}
Both the aim and the tracking speed are affected by multiscale stochastic volatility in the representation above.
\end{remark}

\section{Examples and Numerical Solutions}\label{sec:numeric}

We present numerical examples to demonstrate that the asymptotic approximation can be computed efficiently under a variety of practically relevant models. We then show how the proposed algorithms improve Profit \& Loss (PnL) through Monte Carlo simulations.

Each example highlights how the speed and accuracy of execution evolve when incorporating the asymptotic corrections. Our primary goal is to evaluate how PnL differs when executing trades at a constant rate versus the corrected, optimal trading rate.

\subsection{Numerical Approximation of Price Impact}

To implement the Monte Carlo simulation, particularly in the slow-scale regime, we employ the approximation method described in Appendix \ref{ap:derivatives}. Solving the full nonlinear system at each time step and for every sample path would be computationally prohibitive. Therefore, we apply the first-order approximation introduced in Section \ref{sec:approx} to compute all the coefficients \(A\)'s.

\begin{figure}[htbp]
    \centering
    \includegraphics[width=1.\linewidth]{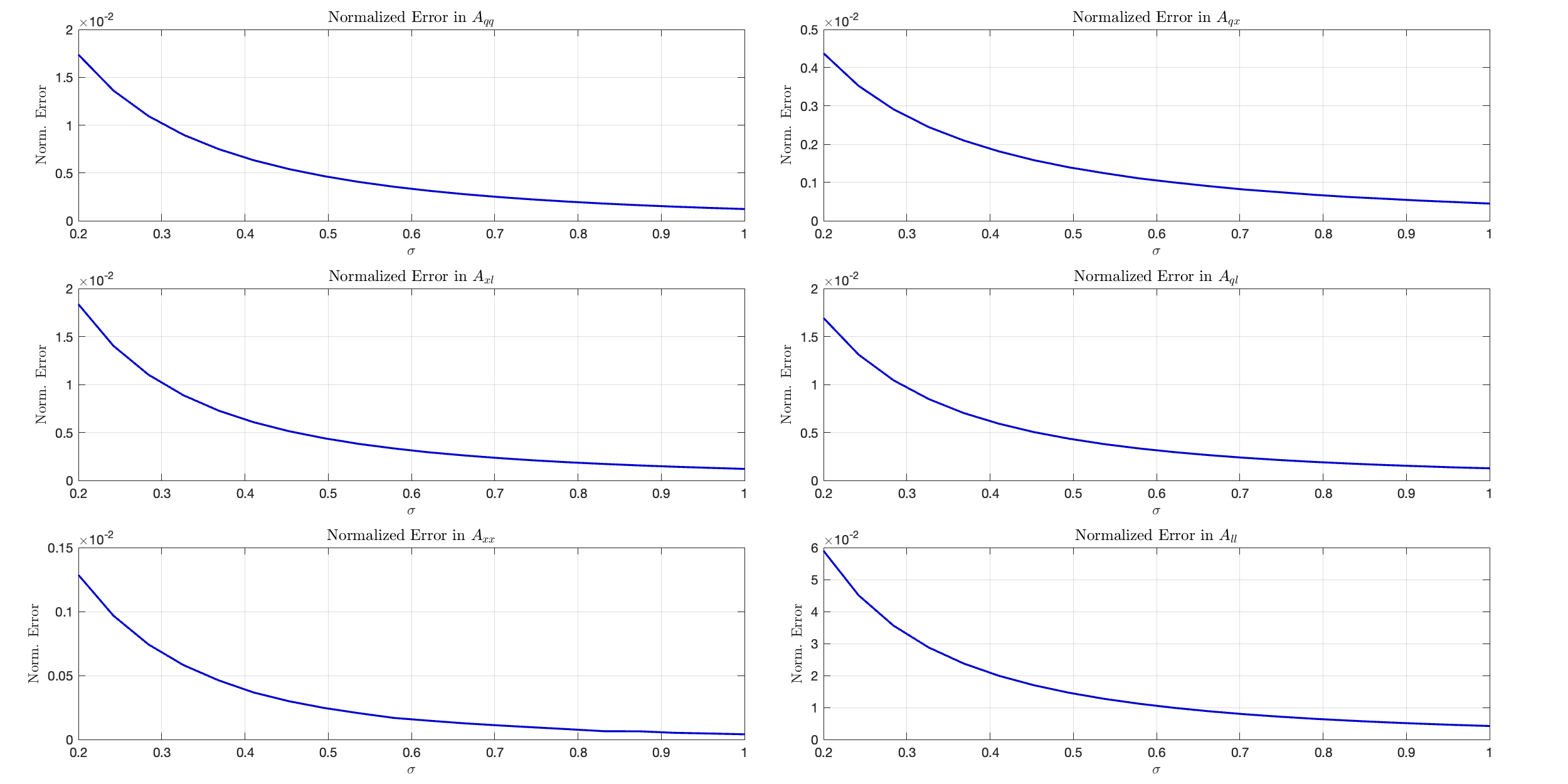}
    \caption{Normalized error of each coefficient using the first-order small price impact approximation from Section~\ref{sec:approx}. The parameter values used are: \(\rho = 0.2\), \(\gamma = 5\), \(K = 1\), \(\lambda = \beta = 1\), \(\theta = 0.1\), \(\kappa = 1\), and \(\eta = 1\).}
    \label{fig:coeff_approx}
\end{figure}

We compare the approximated coefficients to the exact solutions of the system and calculate the normalized error across a wide range of volatility values \(\sigma\), as shown in Figure~\ref{fig:coeff_approx}.

\begin{figure}[htbp]
    \centering
    \includegraphics[width=1\linewidth]{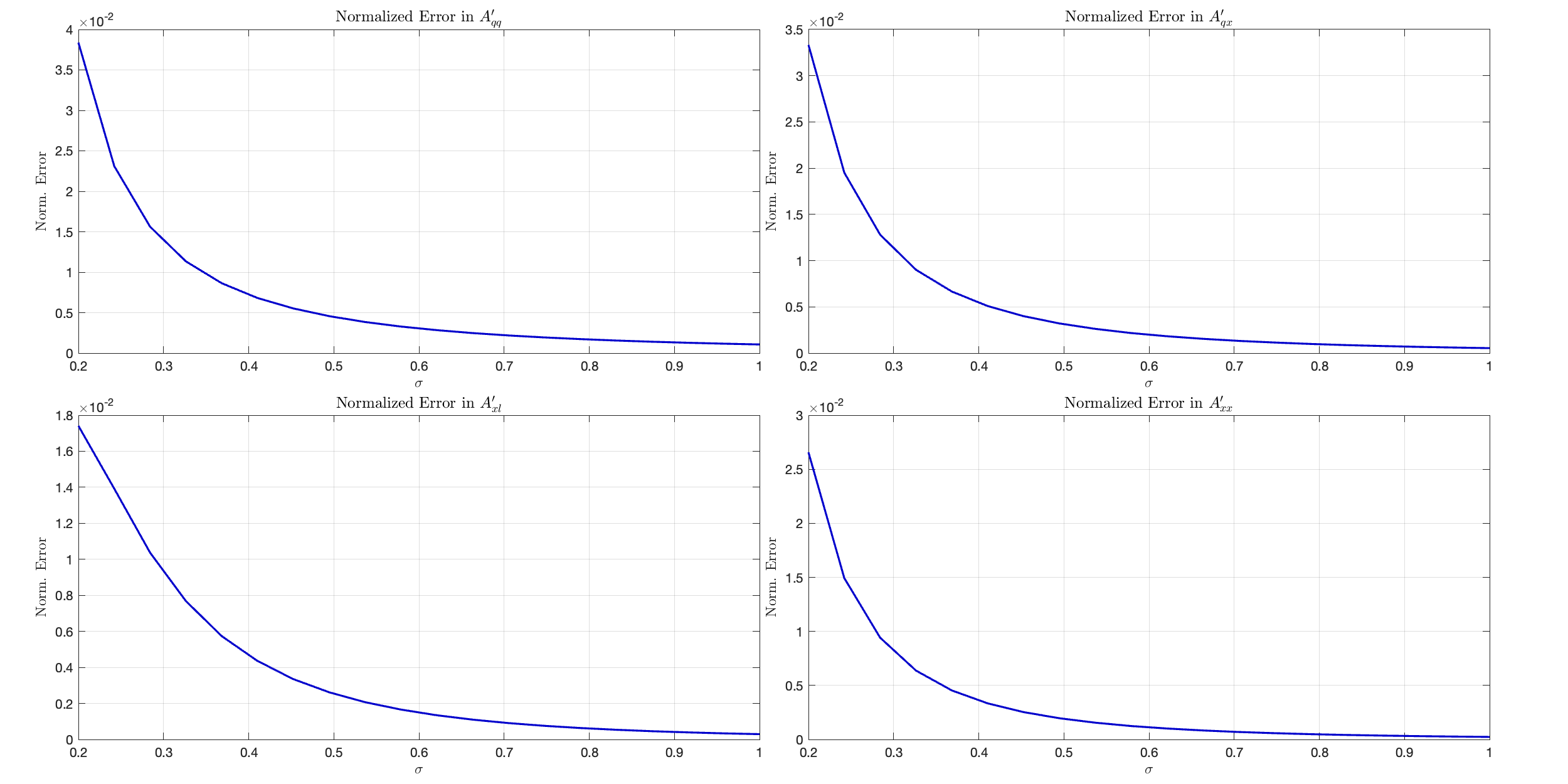}
    \caption{Normalized error of the key derivatives used in computing the \(B\) terms. Parameters are the same as in Figure~\ref{fig:coeff_approx}.}
    \label{fig:der_approx}
\end{figure}

We also compute the normalized error for the four main derivatives of the coefficients, which are necessary to evaluate the \(B_q\) and \(B_l\) terms in the first-order correction. These derivative approximations, based on Appendix~\ref{ap:derivatives}, are illustrated in Figure~\ref{fig:der_approx}.

In both figures, the errors are generally of order \(\mathcal{O}(\theta^2)\), validating the use of the first-order approximation in the Monte Carlo simulation to reduce computational cost. Moreover, we observe that the normalized error decreases rapidly as the volatility \(\sigma\) increases, consistently across all coefficients and their derivatives.

\subsection{Fast Factor Example}\label{subsec:fast_ex}

We assume that the dynamics of the asset price and its stochastic volatility follow Heston’s model. Specifically, as in Example~1 in Section~\ref{subsec:examples}, we have
\[
\sigma(y) = \sqrt{y}, \quad b(y) = \chi(\mu - y), \quad \text{and} \quad a(y) = \psi\sqrt{y},
\]
where \(\chi\), \(\psi\), and \(\mu\) are constant parameters. We assume the standard Feller condition \(\psi^2 < 2\mu\), which guarantees positivity of the volatility process and depends only on constants, independent of time scaling.

In this setting, the optimal trading strategy, including the second-order correction term discussed in Section~\ref{sec:fast} is given by:
\begin{align*}
    u^*(q,l,x,y) =\,& \frac{1}{K} \left((\lambda - A_{qq} + \lambda \, A_{ql}) \, q + (A_{ql} + \lambda \, A_{ll}) \, l + (A_{qx} + \lambda \, A_{xl}) \, x\right) 
    - \frac{1}{K} \, \varepsilon \, \frac{\gamma}{\theta}(y - \mu) \, q.
\end{align*}

\begin{figure}[htbp]
    \centering
    \includegraphics[width=1\linewidth]{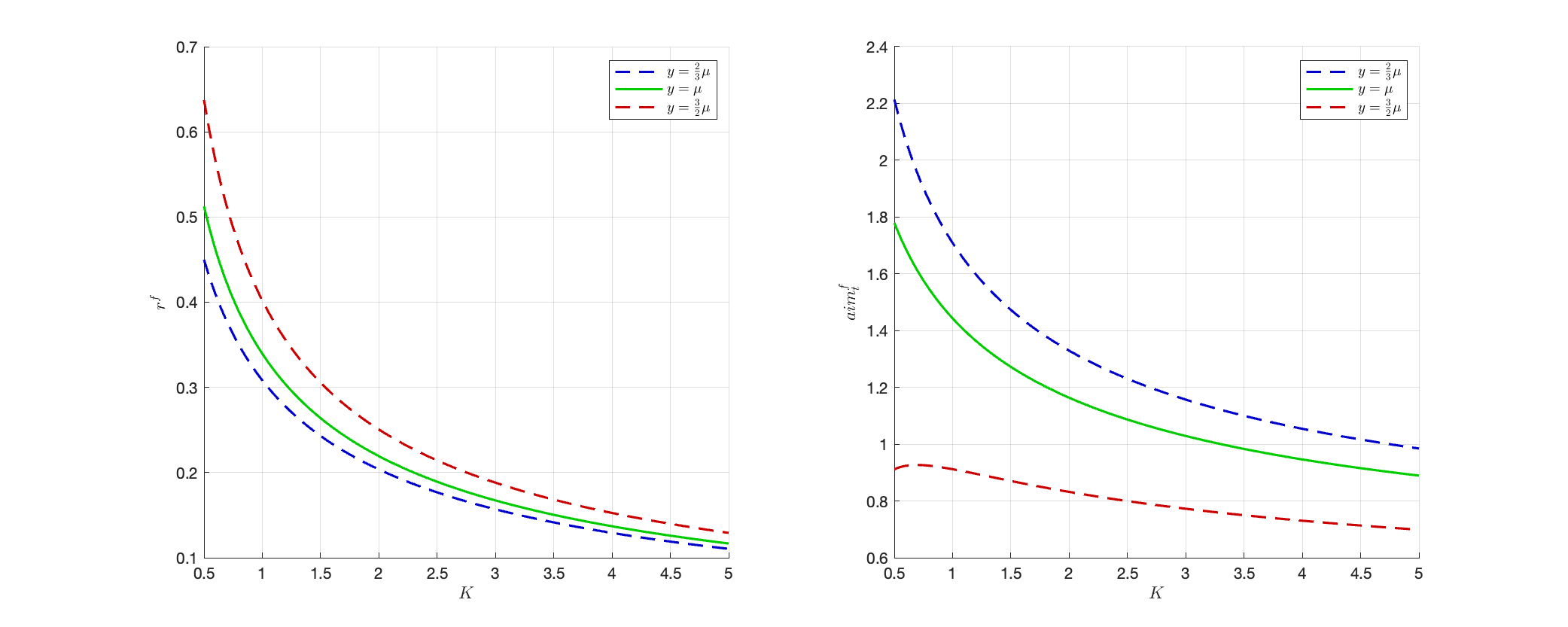}
    \caption{Optimal trading speed and aim portfolio for varying values of the transaction cost parameter \(K\), with and without the second-order correction term. We use the following numerical values for the parameters: \(\gamma = 5\), \(\lambda = \beta = 0.1\), \(\kappa = \eta = 1\), \(\rho_1 = 0.5\), \(K = 1\), \(x = 1\), \(\mu = 0.2\), \(\chi = 1\), \(\psi = 0.25\), and \(\varepsilon = 0.25\).}
    \label{fig:speed_aim_fast}
\end{figure}

We illustrate the trading speed and the corresponding aim portfolio under this strategy. Figure~\ref{fig:speed_aim_fast} shows how these quantities vary as a function of the transaction cost parameter \(K\). We highlight that the solid line, corresponding to the case \(y = \mu\), represents the speed without the correction term, denoted by \(r^c\). As noted earlier, when the volatility exceeds its long-term average \(\mu\), the correction term suggests increasing the adjusted trading speed \(r^f\). Conversely, when the volatility falls below \(\mu\), the correction acts to reduce the trading speed. We also observe the opposite effect for the targeted tracking portfolio \(aim^f\): it decreases when the volatility is above the average \(\mu\), and increases when the volatility is below \(\mu\).

\begin{figure}[hbp]
    \centering
    \includegraphics[width=1\linewidth]{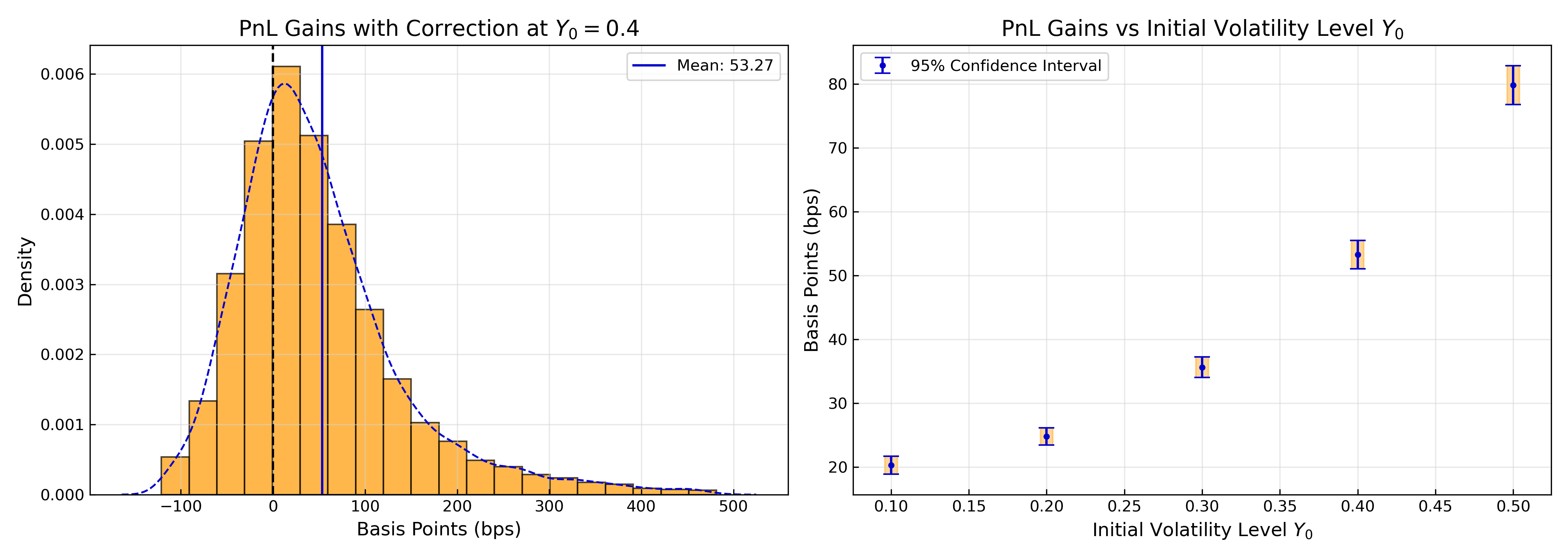}
    \caption{Monte Carlo simulation results: comparison of portfolio PnL under the constant volatility case strategy and the corrected strategy. Parameter's values are as in Figure~\ref{fig:speed_aim_fast} with \(\gamma = 2\), \(q_0=0\), \(l_0 = 0\), \(x_0=5\).}
    \label{fig:mc_ff_pnl}
\end{figure}

With a time horizon of \(T = 2\) years, we use Monte Carlo simulation to compare the realized PnL of two strategies: (i) the constant-rate optimal strategy, and (ii) the corrected strategy that incorporates fast mean-reverting stochastic volatility. The results are shown in Figure~\ref{fig:mc_ff_pnl}.

As observed in Figure~\ref{fig:mc_ff_pnl}, the second-order correction has a significant impact on portfolio performance. In particular, the corrected strategy leads to a substantial improvement in PnL, with an average increase of 53.27 basis points relative to the constant volatility strategy, when \(Y_0 = 0.4\). 

Moreover, we compute the variance of the PnL in both cases: under the optimal strategy assuming constant volatility and under the corrected strategy. In particular, we observe that the variance is lower in the corrected case, namely \(0.6286\) compared to \(0.6515\) without the correction.


\subsection{Slow Factor Example}

\citet{kraft2005optimal} considered a one-factor stochastic volatility model in which the volatility factor \(Z_t\) follows a Cox-Ingersoll-Ross (CIR) process, with dynamics defined by
\[\sigma(z) = \sqrt{z}, \quad c(z) = m - z,\quad \text{and} \quad g(z) = \beta\sqrt{z}.
\]
We again assume the standard Feller condition \(\beta^2 < 2m\), which notably does not depend on the time scale parameter \(\delta\).

\begin{figure}[htbp]
    \centering
    \includegraphics[width=1\linewidth]{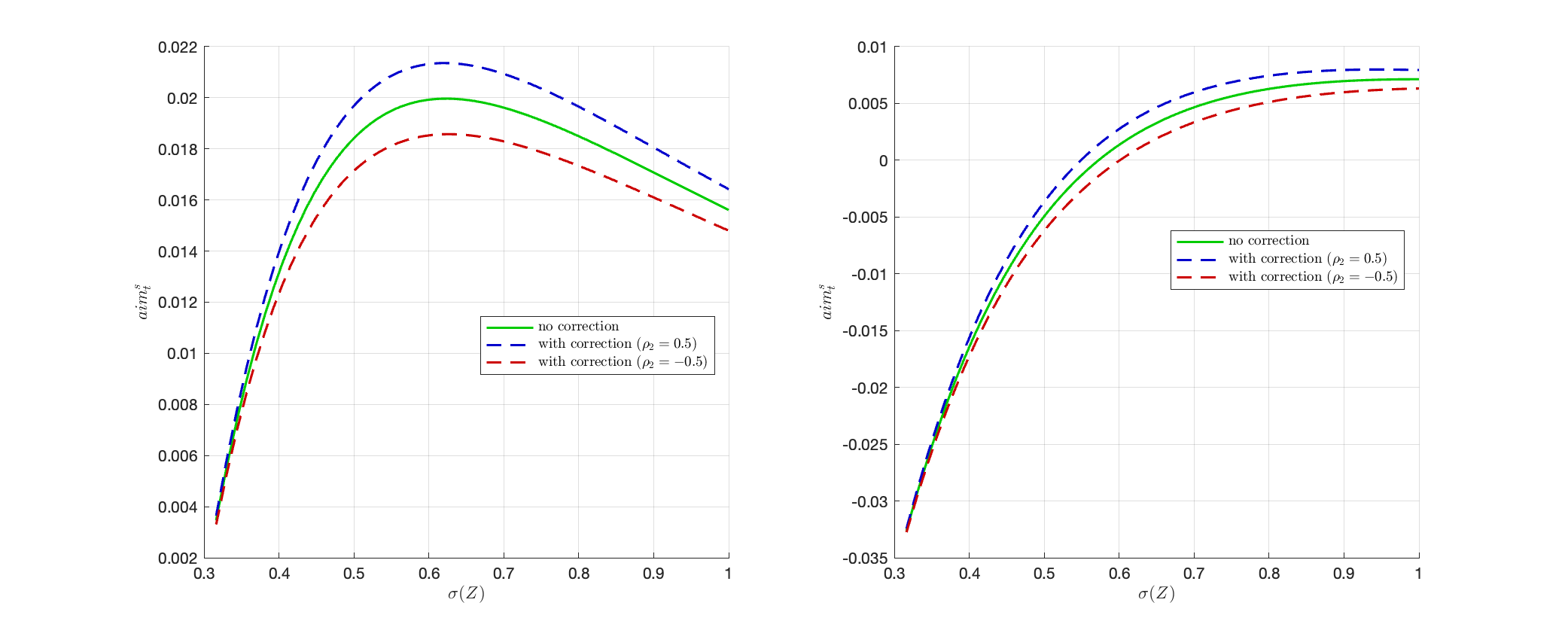}
    \caption{Comparison of the aim portfolio under the constant and corrected strategies, across a range of volatility values \(\sigma(z)\). The left plot is consturcted choosing \(x=0.2, l=0\), and the right plot with \(x=0.2, l=0.05\). We use the parameter values \(\gamma = 5\), \(\lambda = \beta = 0.1\), \(\kappa = \eta = 1\), \(\rho_2 = 0.5\), \(K = 1\), \(m = 0.2\), \(\beta = 0.25\), and \(\delta = 0.5\).} 
    \label{fig:aim_slow}
\end{figure}

We provide plots to compare the aim portfolio under the constant volatility strategy versus the corrected strategy. As shown in Section~\ref{sec:slow}, the trading speed remains unchanged in this regime. In particular, Figure~\ref{fig:aim_slow} illustrates how the optimal target portfolio evolves with respect to the variables \(x\) and \(l\), across a range of values for the volatility factor \(z\). The plots confirm the earlier observation regarding the effect of the correlation between the Brownian motion driving expected returns and the volatility factor \(Z\). Specifically, when this correlation is positive, the targeted portfolio \(aim^s_t\) is larger compared to the constant volatility case. The opposite holds in the case of negative correlation, where the corrected aim is reduced relative to the uncorrected benchmark.

\begin{figure}[htbp]
    \centering
    \includegraphics[width=1\linewidth]{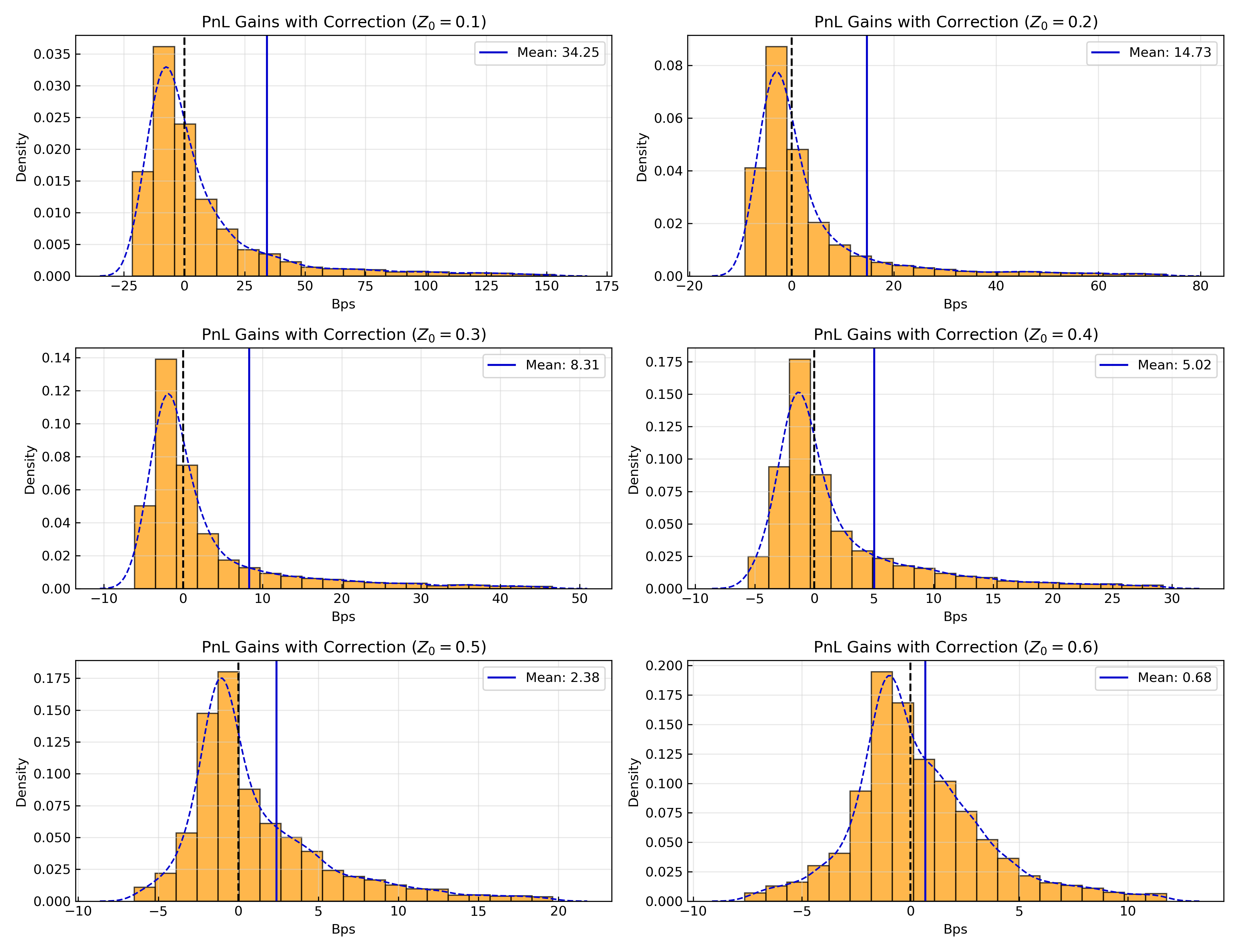}
    \caption{Monte Carlo simulation: empirical PDFs of PnL gains under the corrected strategy, for different initial values \(Z_0\) of the volatility factor. Parameters are used as in Figure~\ref{fig:aim_slow}, with \(q_0=0\), \(l_0=0\), \(x_0 = 2\) and \(\delta = 0.25^2\).}
    \label{fig:mc_slow}
\end{figure}

Using a time horizon of \(T = 2\) years, we perform Monte Carlo simulations to compare the realized portfolio PnL under the constant-rate strategy and the corrected strategy. Figure~\ref{fig:mc_slow} shows the empirical probability density functions (PDFs) of the PnL gains for various initial values of the volatility factor \(z\). The corresponding statistical summaries are provided in Table~\ref{tab:stats}. Notably, even though the time scale parameter \(\delta\) in this slow-factor setting is much smaller than the \(\varepsilon\) used in the fast-scale example in Section~\ref{subsec:fast_ex}, the PnL gains are of comparable magnitude. This supports our earlier conclusion that the slow factor has a more significant effect in correcting the constant volatility strategy.

\vspace{2cm}

\clearpage

\begin{table}[htp]
    \centering
    \caption{Statistical summary of PnL gains (in basis points) under the corrected strategy for various initial values of the volatility factor \(Z_0\). Parameters are used as in Figure~\ref{fig:mc_slow}.}
    \label{tab:stats}
    \begin{tabular}{lccc}
        \toprule
        $Z_0$ & Mean (bps) & Std (bps) & 95\% Confidence Interval \\
        \midrule
        0.1 & 34.2532 & 235.7638 & [29.6322, 38.8741] \\
        0.2 & 14.7266 & 79.2333  & [13.1736, 16.2796] \\
        0.3 & 8.3127  & 39.4801  & [7.5389, 9.0865] \\
        0.4 & 5.0165  & 24.3310  & [4.5396, 5.4934] \\
        0.5 & 2.3825  & 15.0484  & [2.0875, 2.6774] \\
        0.6 & 0.6804  & 10.7493  & [0.4698, 0.8911] \\
        \bottomrule
    \end{tabular}
\end{table}

\section{Conclusion}\label{sec:conclusion}

Empirical studies have shown that stochastic volatility can be effectively modeled using two factors, one fast and one slow, as proposed by \citet{fouque2011multiscale}. In our analysis, we extend the classical optimal execution problem under constant volatility, as considered by \citet{garleanu2013dynamic}, by incorporating this multiscale volatility framework.

The impact of stochastic volatility on optimal execution can be systematically studied using asymptotic approximations, which preserve the tractability of the resulting optimal strategy. In particular, we derive the first two terms of the asymptotic expansion of the \citet{garleanu2013dynamic} value function, separately in the cases of a single fast and a single slow volatility factor. Moreover, in Section~\ref{sec:multi}, we demonstrate how these approximations can be combined in the general case of multiscale stochastic volatility.

Rigorous mathematical analysis establishing the accuracy of such asymptotic approximations has been carried out for the option pricing problem in \citet{fouque2011multiscale}, and more recently for the classical Merton problem in \citet{fouque2021}. In the context of our problem setting, a similar theoretical justification is left for future work.

We have also proposed a method using small parameter approximations to solve the system derived in \cite{garleanu2016} for relatively small price impact. Through numerical examples and Monte Carlo simulations, we have demonstrated that the resulting optimal strategies indeed lead to more efficient portfolio management and improvements in Profit and Loss when volatility exhibits distinct time scale fluctuations.

To extend this work, there are several directions that can be considered to further enrich the model and capture more complex behaviors, thereby achieving a more accurate representation of real-world market dynamics.

\begin{enumerate}
    \item \textbf{Stochastic Liquidity.} The impact of stochastic liquidity on the optimal portfolio, first formulated by \citet{almgren2012optimal} in continuous time, remains both relevant and challenging. In fact, we observe that the constant parameter \(\lambda\) in our current setup varies over the course of the trading day, typically peaking at the market open and close. A more realistic future extension would model \(\lambda\) as a time-varying process, potentially with joint dynamics alongside other sources of stochasticity in the market.

    \item \textbf{Nonlinear Price Impact.} In our analysis, we adopt a simple linear price impact model proportional to the trading rate. However, the literature contains more sophisticated models involving general propagator functions to better reflect the intrinsic characteristics of market impact. That said, enriching the model often comes at the cost of losing tractability in the resulting optimal strategy, and striking a balance between realism and analytical feasibility remains a key consideration.

    \item \textbf{Partial Information and Expert Opinions.} In this paper, we assume that the predictable return is fully observable and modeled as an Ornstein–Uhlenbeck process. However, in practice, signals are often observed with a high signal-to-noise ratio, making it important to assess the impact of partial information on optimal trading behavior. The problem of optimal trading under partial observation and expert opinions has been analyzed in \citet{Fouque2017perturbation} and the references therein, as well as in the more recent work by \citet{MuhleKarbe2023dynamic}.

\end{enumerate}

\appendix

\section{Second Order Small Price Impact approximation}\label{appendix:pi}

In this appendix, we provide the complete second-order for the small price impact approximation introduced in Section \ref{sec:model}.  
At order $\theta^{2}$, the second order correction terms for all the coefficients can be explicitly computed. These results are presented below:

\begin{align*}
    A_{qq}^{(2)} = \frac{M_{qq}}{N_{qq}}, \quad A_{ll}^{(2)} = \frac{M_{ll}}{N_{ll}}, \quad A_{xx}^{(2)} = \frac{M_{xx}}{N_{xx}}, \quad A_{qx}^{(2)} = \frac{M_{qx}}{N_{qx}}, \\
    A_{xl}^{(2)} = \frac{M_{xl}}{N_{xl}}, \quad A_{ql}^{(2)} = \frac{M_{ql}}{N_{ql}}, \quad A_{0}^{(2)} = \frac{M_{0}}{N_{0}},
\end{align*}
where we have that,
\begin{align*}
M_{qq} &= - {A_{qq}^{(1)}}^2 + 2 \, A_{qq}^{(1)} \, \lambda - \lambda^2 + 2 \, A_{ql}^{(1)} \, A_{qq}^{(0)} \, \lambda,\\
N_{qq} &= 2 \, A_{qq}^{(0)} + K \, \rho,\\
M_{ll} &= 2 \, {A_{ql}^{(1)}}^2 \, {A_{qq}^{(0)}}^2 + 3 \, {A_{ql}^{(1)}}^2 \, A_{qq}^{(0)} \, K \, \rho + {A_{ql}^{(1)}}^2 \, K^2 \, \rho^2,\\
N_{ll} &= K \, \rho \, (2 \, A_{qq}^{(0)} + K \, \rho) \, (A_{qq}^{(0)} + K \, \rho).
\end{align*}

\begin{align*}
M_{xx} &= 2 \, {A_{qq}^{(0)}}^2 \, {A_{qx}^{(1)}}^2 - 4 \, A_{qq}^{(0)} \, A_{qq}^{(1)} \, A_{qx}^{(0)} \, A_{qx}^{(1)} + 4 \, A_{qq}^{(0)} \, A_{qx}^{(0)} \, A_{qx}^{(1)} \, \lambda \\
&\quad + 4 \, A_{xl}^{(1)} \, A_{qq}^{(0)} \, A_{qx}^{(0)} \, K \, \lambda \, \rho + 4 \, A_{xl}^{(1)} \, \kappa \, A_{qq}^{(0)} \, A_{qx}^{(0)} \, K \, \lambda + 3 \, A_{qq}^{(0)} \, {A_{qx}^{(1)}}^2 \, K \, \rho \\
&\quad + 2 \, \kappa \, A_{qq}^{(0)} \, {A_{qx}^{(1)}}^2 \, K + 2 \, {A_{qq}^{(1)}}^2 \, {A_{qx}^{(0)}}^2 - 4 \, A_{qq}^{(1)} \, {A_{qx}^{(0)}}^2 \, \lambda - 2 \, A_{qq}^{(1)} \, A_{qx}^{(0)} \, A_{qx}^{(1)} \, K \, \rho \\
&\quad + 2 \, A_{ql}^{(1)} \, {A_{qx}^{(0)}}^2 \, K \, \lambda \, \rho + 2 \, {A_{qx}^{(0)}}^2 \, \lambda^2 + 2 \, A_{qx}^{(0)} \, A_{qx}^{(1)} \, K \, \lambda \, \rho + 2 \, A_{xl}^{(1)} \, A_{qx}^{(0)} \, K^2 \, \lambda \, \rho^2 \\
&\quad + 2 \, A_{xl}^{(1)} \, \kappa \, A_{qx}^{(0)} \, K^2 \, \lambda \, \rho + {A_{qx}^{(1)}}^2 \, K^2 \, \rho^2 + \kappa \, {A_{qx}^{(1)}}^2 \, K^2 \, \rho,\\
N_{xx} &= K \, (2 \, A_{qq}^{(0)} + K \, \rho) \, (2 \, \kappa + \rho) \, (A_{qq}^{(0)} + K \, \kappa + K \, \rho),\\
M_{qx} &= {A_{qq}^{(1)}}^2 \, A_{qx}^{(0)} + A_{qx}^{(0)} \, \lambda^2 - 2 \, A_{qq}^{(0)} \, A_{qq}^{(1)} \, A_{qx}^{(1)} + 2 \, A_{qq}^{(0)} \, A_{qx}^{(1)} \, \lambda \\
&\quad - 2 \, A_{qq}^{(1)} \, A_{qx}^{(0)} \, \lambda - 2 \, {A_{qq}^{(0)}}^2 \, A_{xl}^{(1)} \, \lambda - A_{qq}^{(1)} \, A_{qx}^{(1)} \, K \, \rho \\
&\quad + A_{qx}^{(1)} \, K \, \lambda \, \rho + A_{ql}^{(1)} \, A_{qx}^{(0)} \, K \, \lambda \, \rho - A_{qq}^{(0)} \, A_{xl}^{(1)} \, K \, \lambda \, \rho,\\
N_{qx} &= (2 \, A_{qq}^{(0)} + K \, \rho) \, (A_{qq}^{(0)} + K \, \kappa + K \, \rho),
\end{align*}

\begin{align*}
M_{ql} &= 2 \, A_{ql}^{(1)} \, A_{qq}^{(0)} \, A_{qq}^{(1)} - 2 \, A_{ql}^{(1)} \, A_{qq}^{(0)} \, \lambda + A_{ql}^{(1)} \, K^2 \, \beta \, \rho + 2 \, A_{ql}^{(1)} \, A_{qq}^{(0)} \, K \, \beta \\
&\quad + A_{ql}^{(1)} \, A_{qq}^{(1)} \, K \, \rho - A_{ql}^{(1)} \, K \, \lambda \, \rho,\\
N_{ql} &= (2 \, A_{qq}^{(0)} + K \, \rho) \, (A_{qq}^{(0)} + K \, \rho),\\
M_0 &= \eta \Big(2 \, {A_{qq}^{(0)}}^2 \, {A_{qx}^{(1)}}^2 - 4 \, A_{qq}^{(0)} \, A_{qq}^{(1)} \, A_{qx}^{(0)} \, A_{qx}^{(1)} + 4 \, A_{qq}^{(0)} \, A_{qx}^{(0)} \, A_{qx}^{(1)} \, \lambda \\
&\quad + 4 \, A_{xl}^{(1)} \, A_{qq}^{(0)} \, A_{qx}^{(0)} \, K \, \lambda \, \rho + 4 \, A_{xl}^{(1)} \, \kappa \, A_{qq}^{(0)} \, A_{qx}^{(0)} \, K \, \lambda + 3 \, A_{qq}^{(0)} \, {A_{qx}^{(1)}}^2 \, K \, \rho \Big),\\
N_0 &= 2 \, K \, \rho \, (2 \, A_{qq}^{(0)} + K \, \rho) \, (2 \, \kappa + \rho) \, (A_{qq}^{(0)} + K \, \kappa + K \, \rho).
\end{align*}

\section{Coefficient's Derivatives Approximation}\label{ap:derivatives}

To compute the correction terms in the slow scale stochastic volatility model, we first compute the derivatives of the coefficients.

We consider the initial system consisting of equations (\ref{eq: eq1})–(\ref{eq: eq7}). In the slow scale volatility case, the coefficients depend on \(z\), since the volatility \(\sigma^2\) is a function of \(z\). Taking the derivative of all those equations with respect to \(z\), we obtain the following system:
{\scriptsize
\begin{align*}
    \tfrac{\rho}{2} A_{qq}^{\prime}(z) 
    - \tfrac{1}{2} \gamma \sigma(z) \sigma'(z) 
    - \tfrac{1}{K} (\lambda - A_{qq}(z) + \lambda A_{ql}(z)) (-A_{qq}^{\prime}(z) + \lambda A_{ql}^{\prime}(z)) &= 0, \\
    \tfrac{1}{K} \left( (A_{ql}(z) + \lambda A_{ll}(z)) (-A_{qq}^{\prime}(z) + \lambda A_{ql}^{\prime}(z)) 
    + (A_{ql}^{\prime}(z) + \lambda A_{ll}^{\prime}(z)) (\lambda - A_{qq}(z) + \lambda A_{ql}(z)) \right) 
    - (\beta + \rho) A_{ql}^{\prime}(z) &= 0, \\
    \tfrac{1}{K} \left( (A_{qx}(z) + \lambda A_{xl}(z)) (-A_{qq}^{\prime}(z) + \lambda A_{ql}^{\prime}(z)) 
    + (A_{qx}^{\prime}(z) + \lambda A_{xl}^{\prime}(z)) (\lambda - A_{qq}(z) + \lambda A_{ql}(z)) \right) 
    - (\rho + \kappa) A_{qx}^{\prime}(z) &= 0, \\
    \tfrac{1}{K} (A_{ql}(z) + \lambda A_{ll}(z)) (A_{ql}^{\prime}(z) + \lambda A_{ll}^{\prime}(z)) 
    - \left( \tfrac{\rho}{2} + \beta \right) A_{ll}^{\prime}(z) &= 0, \\
    \tfrac{1}{K} \left( (A_{ql}(z) + \lambda A_{ll}(z)) (A_{qx}^{\prime}(z) + \lambda A_{xl}^{\prime}(z)) 
    + (A_{ql}^{\prime}(z) + \lambda A_{ll}^{\prime}(z)) (A_{qx}(z) + \lambda A_{xl}(z)) \right) 
    - (\rho + \beta + \kappa) A_{xl}^{\prime}(z) &= 0, \\
    \tfrac{1}{K} (A_{qx}(z) + \lambda A_{xl}(z)) (A_{qx}^{\prime}(z) + \lambda A_{xl}^{\prime}(z)) 
    - \left( \tfrac{\rho}{2} + \kappa \right) A_{xx}^{\prime}(z) &= 0, \\
    \tfrac{\eta}{2} A_{xx}^{\prime}(z) - \rho A_0^{\prime}(z) &= 0.
\end{align*}
}
Should the values of the coefficients are known, the system above is linear in the derivatives of the coefficients. By using the previous approximation of the coefficients from the price impact asymptotic expansion and solving this linear system, we obtain an approximation of the derivatives to order \(\mathcal{O}(\theta^2)\), as computed in the specific example in Section~\ref{sec:numeric}.

The system above can be rewritten in the following matrix form:
\[
    M(z) \cdot \bar{a}(z) = b(z),
\]
where the matrix \(M(z)\) is given below (with all coefficients \(A\) depending on \(z\)):
{\tiny
\[
\left[
\begin{array}{c@{}c@{}c@{}c@{}c@{}c@{}c}
\begin{array}{c}
     \tfrac{1}{K} (\lambda - A_{qq} + \lambda A_{ql}) \\
     + \tfrac{\rho}{2}
\end{array} & 
-\tfrac{\lambda}{K} (\lambda - A_{qq} + \lambda A_{ql}) & 
0 & 0 & 0 & 0 & 0 \\[1em]

-\tfrac{1}{K} (A_{ql} + \lambda A_{ll}) & 
\begin{array}{c}
     \tfrac{1}{K} (\lambda - A_{qq} + \lambda A_{ql}) \\
     -(\beta + \rho)
\end{array} & 
0 & \tfrac{\lambda}{K} (A_{ql} + \lambda A_{ll}) & 
0 & 0 & 0 \\[1em]

\tfrac{1}{K} (A_{qx} + \lambda A_{xl}) & 
\tfrac{\lambda}{K} (A_{qx} + \lambda A_{xl}) & 
\begin{array}{c}
     (\rho + \kappa) \\
     - \tfrac{1}{K} (\lambda - A_{qq} + \lambda A_{ql})
\end{array} & 
0 & -\tfrac{\lambda}{K} (A_{qx} + \lambda A_{xl}) & 
0 & 0 \\[1em]

0 & \tfrac{1}{K}(A_{ql} + \lambda A_{ll}) & 
0 & 
\begin{array}{c}
     \tfrac{\lambda}{K} (A_{ql} + \lambda A_{ll}) \\
     -(\tfrac{\rho}{2} + \beta)
\end{array} & 
0 & 0 & 0 \\[1em]

0 &\tfrac{1}{K}(A_{qx} + \lambda A_{xl}) & 
\tfrac{1}{K}(A_{ql} + \lambda A_{ll}) & \tfrac{\lambda}{K} (A_{qx} + \lambda A_{xl}) & 
\begin{array}{c}
     \tfrac{\lambda}{K} (A_{ql} + \lambda A_{ll}) \\
     - (\rho + \beta + \kappa)
\end{array} & 
0 & 0 \\[1em]

0 & 0 & \tfrac{1}{K}(A_{qx} + \lambda A_{xl}) & 0 & \tfrac{\lambda}{K}(A_{qx} + \lambda A_{xl}) &  -\left(\tfrac{\rho}{2} + \kappa\right) & 
0 \\[1em]

0 & 0 & 0 & 0 & 0 & \tfrac{\eta}{2} & -\rho
\end{array}
\right],
\]

\[
\bar{a}(z) =
\begin{bmatrix}
A_{qq}^{\prime}(z) \\
A_{ql}^{\prime}(z) \\
A_{qx}^{\prime}(z) \\
A_{ll}^{\prime}(z) \\
A_{xl}^{\prime}(z) \\
A_{xx}^{\prime}(z) \\
A_0^{\prime}(z)
\end{bmatrix}
\quad \text{and} \quad
b(z) =
\begin{bmatrix}
\tfrac{1}{2} \gamma \sigma(z) \sigma'(z) \\
0 \\
0 \\
0 \\
0 \\
0 \\
0
\end{bmatrix}.
\]
}

\section{Second Order Asymptotic for Slow Factor}\label{ap: fast_2}

To solve the PDE \eqref{eq:pde_fast} derived in Section \ref{sec:slow}, we introduced the following asymptotic expansion:
\begin{align*}
    v^{\delta}(q,l,x,z) = v^{(0)}(q,l,x,z) + \sqrt{\delta} \, v^{(1)}(q,l,x,z) + \delta \, v^{(2)}(q,l,x,z) + \cdots.
\end{align*}

We calculated the first-order correction term, which is of order \(\sqrt{\delta}\). We now proceed to compute the second-order correction term, which is of order \(\delta\). Specifically, by substituting the expansion into the PDE and collecting terms of order \(\delta\), we obtain the following PDE:
\[
\mathcal{M}_2 v^{(0)} + \frac{\eta}{2} v^{(2)}_{xx} - \rho v^{(2)} + \frac{(v^{(1)}_q + \lambda v^{(1)}_l)^2}{2K} + \frac{({v^{(2)}}_q + \lambda v^{(2)}_l)(v^{(0)}_q + \lambda q + \lambda v^{(0)}_l)}{K} - \beta l v^{(2)}_l - \kappa x v^{(2)}_x + \mathcal{M}_1 v^{(1)} = 0.
\]

We have already computed the functions \(v_0\) and \(v_1\) in terms of the coefficients \(A_{qq}(z)\), \(A_{xx}(z)\), \(A_{ll}(z)\), \(A_{qx}(z)\), \(A_{xl}(z)\), \(A_{ql}(z)\), \(A_0(z)\), \(B_x(z)\), \(B_q(z)\), and \(B_l(z)\).

This PDE can again be solved using a linear-quadratic ansatz:
\begin{equation*}
v^{(2)}(q,l,x,z) = \frac{1}{2} D_{qq}(z) q^2 + \frac{1}{2} D_{xx}(z) x^2 + \frac{1}{2} D_{ll}(z) l^2 + D_{qx}(z) qx + D_{ql}(z) ql + D_{xl}(z) xl + D_0(z),
\end{equation*}

Substituting this form into the PDE, we match and set to zero the coefficients of the terms \(q^2\), \(x^2\), \(l^2\), \(qx\), \(ql\), \(xl\), and the constant term. Solving the resulting system allows us to determine all the coefficients: \(D_{qq}(z)\), \(D_{xx}(z)\), \(D_{ll}(z)\), \(D_{qx}(z)\), \(D_{ql}(z)\), \(D_{xl}(z)\), and \(D_0(z)\).

The system to solve is given below. Note that it is linear in the coefficients \(D\):

\begin{align*}
    0 &= \frac{\rho}{2} D_{qq}(z) - \mathcal{M}_2 A_{qq}(z) - \frac{(D_{qq}(z) - \lambda D_{ql}(z))(\lambda - A_{qq}(z) + \lambda A_{ql}(z))}{K}, \\
    0 &= 2\mathcal{M}_2 A_{ql}(z) - (\rho + \beta) D_{ql}(z) + \frac{(\lambda D_{ql}(z) + \lambda^2 D_{ll}(z))(\lambda - A_{qq}(z) + \lambda A_{ql}(z))}{K}\\
    &\quad - \frac{(\lambda A_{ll}(z) + A_{ql}(z))(D_{qq}(z) - \lambda D_{ql}(z))}{K}, \\
    0 &= 2\mathcal{M}_2 A_{qx}(z) - (\kappa + \rho) D_{qx}(z) + \frac{(\lambda D_{qx}(z) + \lambda D_{xl}(z))(\lambda - A_{qq}(z) + \lambda A_{ql}(z))}{K}\\
    &\quad - \frac{(\lambda A_{xl}(z) + A_{qx}(z))(D_{qq}(z) - \lambda D_{ql}(z))}{K}, \\
\end{align*}

\begin{align*}
    0 &= \mathcal{M}_2 A_{ll}(z) - \left(\frac{\rho}{2} + \beta \right) D_{ll}(z) + \frac{(\lambda A_{ll}(z) + A_{ql}(z))(\lambda D_{ll}(z) + D_{ql}(z))}{K}, \\
    0 &= 2\mathcal{M}_2 A_{xl}(z) - (\kappa + \rho + \beta) D_{xl}(z) + \frac{(\lambda A_{ll}(z) + A_{ql}(z))(\lambda D_{xl}(z) + D_{qx}(z))}{K}\\
    &\quad + \frac{(\lambda A_{xl}(z) + A_{qx}(z))(\lambda D_{ll}(z) + D_{ql}(z))}{K}, \\
    0 &= \mathcal{M}_2 A_{xx}(z) - (\kappa + \frac{\rho}{2}) D_{xx}(z) + \frac{(\lambda A_{xl}(z) + A_{qx}(z))(\lambda D_{xl}(z) + D_{qx}(z))}{K}, \\
    0 &= 2\mathcal{M}_2 A_{0}(z) + \frac{\eta}{2} D_{xx}(z) - \rho D_0(z) + \frac{(B_q + \lambda B_l)^2}{2K} + \sqrt{\eta} g(z) \rho_2 B_x^\prime(z).
\end{align*}


\end{document}